\documentclass[fleqn]{article}
\usepackage{amsfonts,amssymb, amsmath}
\usepackage{graphicx}
\usepackage{epstopdf}
\usepackage{stfloats}
\usepackage[scriptsize]{caption}
\usepackage{url}
\usepackage{subfig}
\def\noi{\noindent}

\def\Title#1{\noi {{\Large\bf #1}}\\[1ex]}

\def\Aunames#1{\noi{\bf #1}}
\def\auth#1{${}^{#1}$}
\def\Addresses#1{\medskip\noi \protect
    \begin{description}\itemsep -3pt {\small\it #1} \end{description}}
\def\addr#1#2{\item[]{\it #2}}

\def\Abstract#1{\vskip 2mm \begin{center}
        \parbox{16.4cm}{\small\noi #1} \end{center}\medskip}

\def\email#1#2{\footnotetext[#1]{e-mail: #2}\addtocounter{footnote}{1}}

\def\nqq{\hspace*{-2em}}

\def\lal{&&\nqq {}}

\def\beq{\begin{equation}}            
\def\eeq{\end{equation}}              
\def\bear{\begin{eqnarray}}           
\def\bearr{\begin{eqnarray} \lal}     
\def\ear{\end{eqnarray}}              
\def\earn{\nonumber \end{eqnarray}}

\def\dst{\displaystyle}

\def\({\left(}
\def\){\right)}

\def\mn{_{\mu\nu}}

\begin{document}
\twocolumn[

\Title{Testing Lattice Quantum Gravity in 2+1 Dimensions}

\Aunames {Michael K. Sachs\auth{1}
      }

\Addresses{
\addr a {University of California, Davis  \\ One Shields Avenue \\ Davis, CA 95616  } }

\Abstract
  {
 Borrowing techniques from cosmology, I compute the power spectrum of quantum fluctuations in (2+1)-dimensional causal dynamical triangulations, a promising discrete path integral approach to quantum gravity. The results agree with those of canonical quantization to a high degree of precision, providing strong evidence for the equivalence of the two approaches and for the validity of the discrete method. }

] 
\email 1 {mksachs@ucdavis.edu}

\section{Introduction}

Few quantum theories start their lives as quantum theories. Most are born in the familiar classical realm and quantized. There are many ways of performing this quantization, but they mostly fall into two broad categories. The first, path integral quantization, is based on the principal of least action in classical mechanics. An integral is constructed over all field configurations between some initial and final boundary configurations with the action contributing a phase factor to each configuration. In the second method, canonical quantization, variables in the classical theory are promoted to operators on a Hilbert space. The structure of the theory is then defined through commutators of these variables. 

When applying these techniques to general relativity many technical and conceptual difficulties arise. These problems are so severe that no one has yet been able to overcome them except in certain simplified cases. And the situation gets even worse; it is unclear if the results from the different quantization procedures, were we able to carry them out in full, would agree with each other \cite{carlip1}. So the question arises: if there are several theories that yield the same classical behavior but describe very different quantum mechanics, which is the correct quantum theory?

In this paper I will examine the results from two different approaches to quantizing gravity that can be carried out in full. The first is the reduced phase space quantization of (2+1)-dimensional gravity with spherical topology. The reduced dimensionality and simple topology yield a trivial classical spacetime. This simplicity is preserved when the theory is quantized, resulting in a quantum theory with no degrees of freedom. The second approach is causal dynamical triangulations, a lattice regularization of the gravitational path integral. Smooth spacetime is approximated by a triangulated mesh and the path integral is approximated by a sum over inequivalent triangulations. The sum is computed using Monte Carlo techniques, and the resulting spacetime is analyzed using techniques borrowed from the analysis of cosmic microwave background temperature fluctuations. I find that the resulting measurements reveal an extremely spherical spacetime, indicating that the two quantization techniques yield very similar results.

\section{(2+1)-dimensional gravity and canonical quantization} \label{2p1Grav}
The action for general relativity is the Einstein-Hilbert action

\beq										\label{eha}
			S_{EH} =  \frac{1}{16 \pi G } \int_{M} d^{4}x 	\sqrt{-g} \left( R - 2 \Lambda \right),
\eeq

\noindent and the Euler-Lagrange equations give us the Einstein field equations
 
 \beq										\label{efe}
			R\mn - \frac{1}{2} g\mn R +\Lambda g\mn = - 8 \pi G T\mn.
\eeq

As one would expect, working in 2+1 dimensions greatly simplifies the equations of motion. In $d$ dimensions (\ref{efe}) allows $d(d-3)$ local phase space degrees of freedom \cite{carlip2, carlip3}. In $d=4$ this gives us the 4 degrees of freedom that result in gravitational waves. In $d=3$, however, this results in zero local degrees of freedom. Additionally, in 2+1 dimensions the curvature tensor $R_{\mu\nu\rho\sigma}$ depends linearly on the Ricci tensor $R\mn$. This means that vacuum solutions to (\ref{efe}) are flat if $\Lambda = 0$ and of constant curvature if $\Lambda \neq 0$ \cite{carlip2, carlip3}.

So it seems that choosing to work in 2+1 dimensions has given us a fairly boring universe to work with. It turns out that exactly how boring depends on the topology of spacetime. In general our manifolds $M$ will decompose into one-dimensional timelike intervals $I$ and two-dimensional spacelike manifolds $\Sigma$; more compactly $I \times \Sigma$. If we choose $\Sigma$ to be any manifold containing non-contractible curves (e.g., a torus) the resulting holonomies will yield a finite number of global degrees of freedom \cite{carlip3,carlip4}. For our purposes, we want the simplest possible scenario to work with, so we will work with a spacelike manifold in which all curves are contractible. If we choose our constant curvature to be positive, this leaves us with a two-sphere $S^2$. The simplest possible 2+1 dimensional topology we can look at is then $I \times S^2$.

Quantizing even this simple spacetime is far from straightforward. A detailed description of how one goes about doing this can be found in \cite{carlip3}. The result is that if one chooses to apply the classical constraints {\it first} and then quantize the system, one ends up with a quantized spacetime with no local or global degrees of freedom. The Hilbert space is one-dimensional; the 2-sphere that we started with remains a 2-sphere, around which there are no quantum fluctuations. 

\section{Lattice quantum gravity and covariant quantization} \label{LQG}

The basic idea behind quantizing a classical theory using the path integral approach is to form an integral over all possible field configurations between some fixed initial and final configurations. The field configurations we will integrate over are essentially all of the unique geometries characterized by spacetime metrics interpolating between some starting spatial geometry and some ending spatial geometry.  Along with the Einstein-Hilbert action (\ref{eha}), we now have all the ingredients we need to write down a formal path integral for quantum gravity:

\beq										\label{pic}
			Z = \int \mathcal{D} [g] e^{i S_{EH}}.
\eeq

In practice, expressions like (\ref{pic}) are usually calculated using perturbation theory. This approach fails with gravity, however, because the perturbation series is non-renormalizable \cite{gs}. In order to proceed, we need to introduce some kind of regularization scheme. We will do this by approximating the smooth manifolds $M$ by a collection of flat simplices glued together at their edges, very much like a smooth sphere is approximated by a geodesic dome. 

In two dimensions, the simplex used is a triangle. Since the triangle is flat, any curvature in the surface created by glued together triangles is concentrated at vertices $v$ where $i$ triangles come together, and is characterized by the deficit angle: 

\beq										\label{defang}
			\delta_{v} = 2 \pi - \dst\sum_{i} \theta_{vi}.
\eeq

In $n$ dimensions, triangles are replaced with $n$-dimensional simplices connected to other simplices by $(n-2)$-dimensional vertices or ``bones.'' The curvature at each bone is still described by (\ref{defang}), where the dihedral angle $\theta$ is now measured around an $(n-2)$-dimensional bone. The sum over deficit angles, weighted by the ``volume" $V_{b}$ of each bone, is proportional to the integral of the scalar curvature over the surface:

\beq										\label{ndsumtointegral}
			\dst\sum_{b} V_{b} \delta_{b} = \frac{1}{2} \int_{S}  R \sqrt{g} d^{n}x.
\eeq

We are now ready to write down a discrete version of (\ref{eha}), called the Regge action \cite{regge,cms}:

\beq										\label{ra}
			S_{EH} \to S_{R} =  \frac{1}{8 \pi G } \dst\sum_{\mathrm{bones}} V_{b} \delta_{b} - \frac{\Lambda}{8 \pi G }  \dst\sum_{\mathrm{n-simplices}} V_{\mathrm{simplex}}.
\eeq

The path integral (\ref{pic}) over all unique field configurations is then taken to be a sum over all unique triangulations $T$ weighted by $S_{R}$:

\beq										\label{pid}
			Z = \int \mathcal{D} [g] e^{i S_{EH}} \to Z = \dst\sum_{T} \frac{1}{C(T)} e^{i S_{R}},
\eeq

\noindent where $\frac{1}{C(T)}$ is a symmetry factor with $C(T)$ being the order of the automorphism group for triangulation $T$ \cite{cdt1, quant-de-sitter}.

\subsection{Causal dynamical triangulations} \label{CDT}

Historically, discrete approaches to gravity were constructed using Euclidian building blocks. This was mostly done for technical reasons; the weight factors in (\ref{pid}) become real, which is usually necessary for the sums to converge. However, these approaches were unable to reproduce classical spacetimes \cite{cdt1}. Causal dynamical triangulations (CDT) \cite{cdt1} attempts to address this problem by taking the Lorentzian structure of spacetime seriously from the outset. The building blocks are Lorentzian simplices with some (n-1)-dimensional faces being spacelike and some timelike. The action (\ref{ra}) is constructed using these building blocks and then rotated into the Euclidian sector to aid in computation. The resulting construction exhibits three distinct phases. Two of these phases resemble the pathological results from Euclidian dynamical triangulations, a crumpled phase and a branched polymer phase, neither of which seem able to reproduce classical spacetimes. The third phase, however, is characterized by an extended geometry with small fluctuations, as one would hope for from a quantum theory of gravity. This phase is the result of building the causal structure into the theory from the outset, and is the main success of the CDT program. This paper will only focus on the (2+1)-dimensional construction, although the (3+1)-dimensional approach is very similar. Details of both cases can be found in \cite{cdt1}.

\begin{figure}[h]
\centerline{\includegraphics{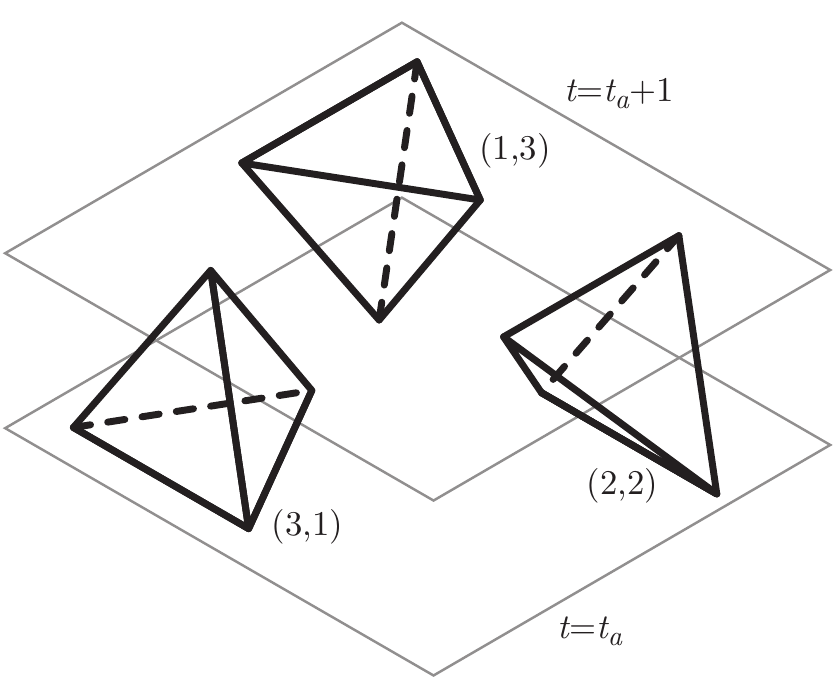}}
\caption{The three simplices of 2+1 CDT.
\label{2p1simplices}}
\end{figure}

In (2+1)-dimensional CDT we use three types of three-dimensional blocks to build our spacetime. Each block is constructed from points that lie in the constant time surface $t=t_{a}$ and the next constant time surface $t=t_{a}+1$. We will label our blocks by ``(number of points at $t_{a}$, number of points at $t_{a}+1$)." With this convention our three blocks are the (3,1) simplex, the (1,3) simplex (a reflection of the (3,1) simplex) and the (2,2) simplex (Figure \ref{2p1simplices}). 

We define the length of edges that lie in constant time surfaces to be $l_{\mathrm {space}}^2 = a^2$ and the length of edges that connect two neighboring timelike surfaces as $l_{\mathrm {time}}^2 = -\alpha a^2$; $\alpha$ is an asymmetry parameter between timelike lengths and spacelike lengths. With these definitions the Regge action (\ref{ra}) in three dimensions becomes \cite{cdt1}:
								
\begin{align}				\label{ra3d}
			S^{(3)}_{R}  = & \frac{1}{8 \pi G} \left( \dst\sum_{ \substack{ \mathrm{spacelike} \\ \mathrm{bones\ } b }}  \frac{1}{i} V_{b} \delta_{b}  +  \dst\sum_{ \substack{ \mathrm{timelike} \\ \mathrm{bones\ } b }} V_{b} \delta_{b} \right) \nonumber \\ &- \frac{\Lambda}{8 \pi G} \left(  \dst\sum_{ \substack{ \mathrm{simplices\ } s }} V_{s} \right).
\end{align}

Following the work in \cite{cdt1}, Lorentzian volumes $V_{i}$ and dihedrial angles $\theta_{i}$ for geometrical elements $i$ are applied to (\ref{ra3d}). Then a series of relationships between all of the bulk variables in the manifold are used, along with the simplifying assumption that $\alpha = 1$\footnote{Even without this assumption, trigonometric identities reduce the action to the form \ref{ra3dBulkEuclSimple}, although the relations \ref{k0k3} become more complicated.} and a choice of units $a = 1$. Lastly the entire action is Wick rotated into the Euclidean sector. The result is \cite{cdt1,3dcdt}:

\beq										\label{ra3dBulkEuclSimple}
			S^{(3)}_{eucl} = -k_0N_0+k_3N_3
\eeq

\noindent with

\begin{align}					\label{k0k3}
			&k_0 = \frac{1}{4G}, \nonumber \\ 
			&k_3= \frac{1}{4 \pi G} \left( 3 \arccos \frac{1}{3} - \pi \right)+ \frac{\Lambda}{48 \pi \sqrt{2}}
\end{align}

\noindent where $N_0$ is the number of vertices in the manifold, $N_3$ is the number of three-simplices, $G$ is Newton's constant, and $\Lambda$ is the cosmological constant. Finally, the discrete path integral (\ref{pid}) becomes:

\beq										\label{pideucl}
			Z = \dst\sum_{T} \frac{1}{C(T)} e^{i S^{(3)}_{R}} \to Z = \dst\sum_{T} \frac{1}{C(T)} e^{- S^{(3)}_{eucl}},
\eeq

\subsection{Monte Carlo Moves} \label{MCM}

\begin{figure}[h]
\centerline{\includegraphics{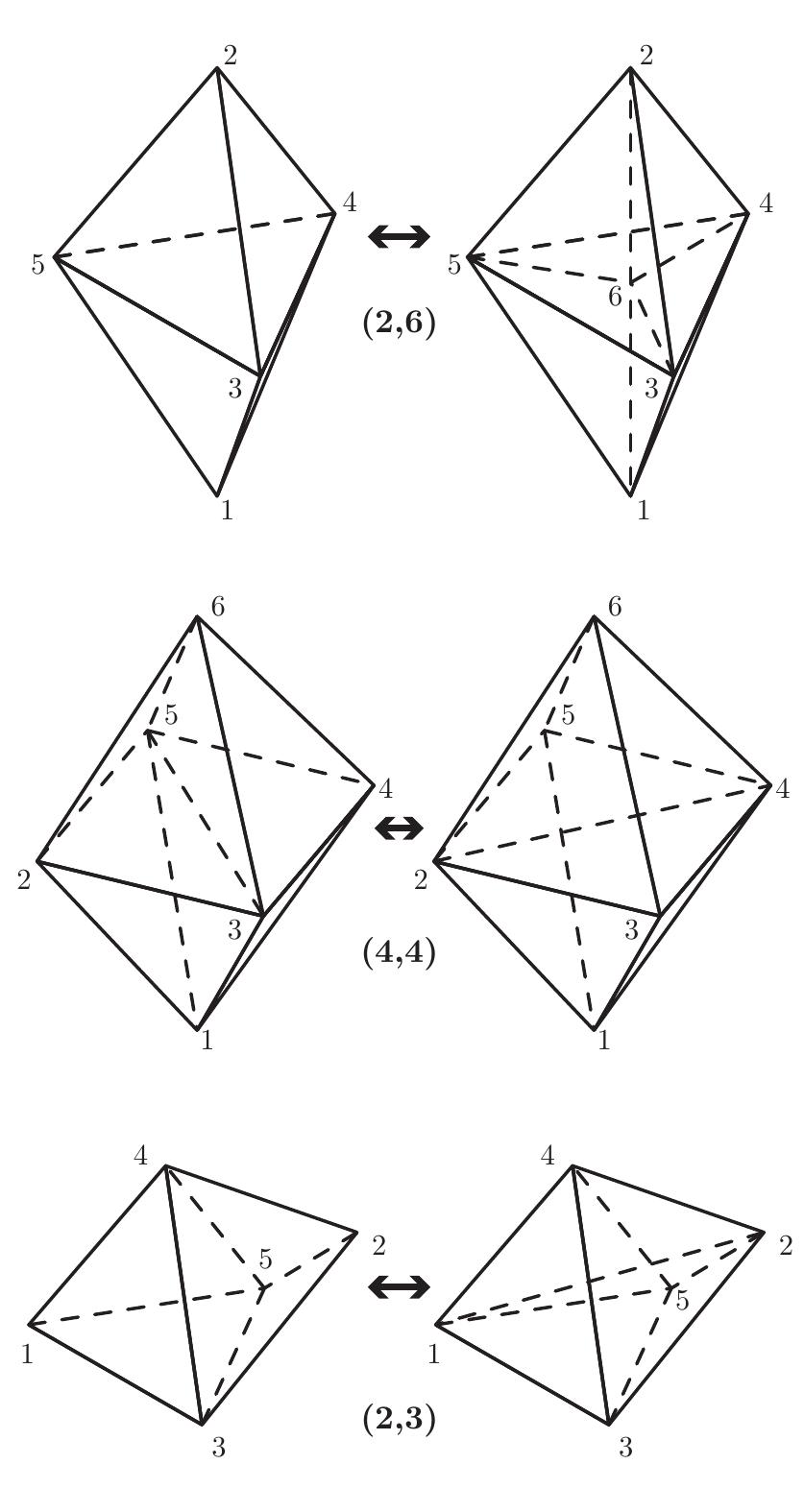}}
\caption{The three Monte Carlo moves that, along with their inverses, make up the five move set (the (4,4) move is it's own inverse).
\label{moves}}
\end{figure}

In order to numerically implement the partition function (\ref{pideucl}), one needs to create a procedure that will explore different allowed triangulations of a given three-dimensional spacetime. This is accomplished by using the Metropolis Monte Carlo algorithm \cite{MetropolisRRTT53} with a series of so-called ``moves," each move being a simple re-triangulation of a set of initial simplices. The constraints on these moves are that they should: (i) preserve the existing time-slicing of the spacetime, (ii) preserve the topology of the spacelike slices and, (iii) be ergodic, that is, any one allowed triangulation can be transformed to any other allowed triangulation by repeated application of these moves. In three dimensions there are five of these moves \cite{cdt1}, four of which are related by inversion symmetry. They are shown in Figure \ref{moves}. Each move is attempted a roughly equal number of times and is accepted or rejected with probabilities depending on how it changes the action (\ref{ra3dBulkEuclSimple}) and how it effects the local geometry. The unit of simulation time is called a sweep and is defined as $N_{3}$ attempted moves (i.e one attempted move for every simplex in the spacetime). The specific CDT implementation that was used for this work is described in \cite{rk}.

\section {Process} \label{P}

The goal of this work is to measure the sphericity of our CDT quantized spacetime, specifically the ensemble average of the sphericity of individual spacelike slices across many sweeps of the simulation. There are four steps required to do this: (i) choose a reference spacelike slice in each sweep to be analyzed, (ii) embed this two-dimensional manifold in three dimensions, (iii) measure the distance from the three-dimensional center of geometry to the embedded manifold and, (iv) analyze the distances across many sweeps using spherical harmonics.

\subsection{Step 1: Choose reference slices} \label{P1}

\begin{figure}[h]
\centerline{\includegraphics{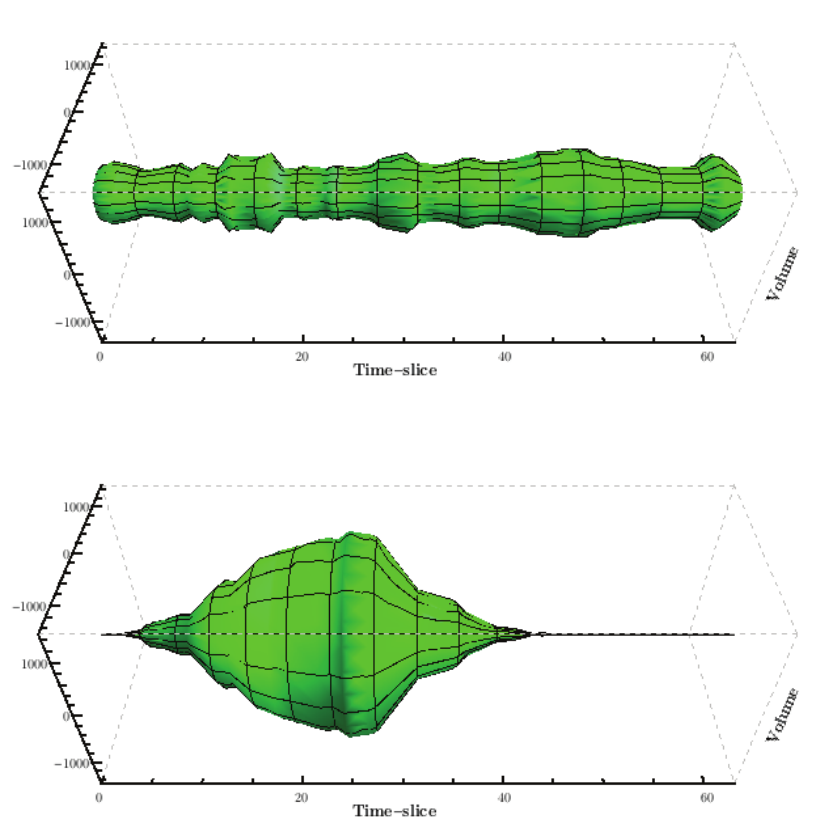}}
\caption{Plots of the volume per time-slice for simulation 3. The top is sweep 500, before the simulation has thermalized. The bottom is sweep 76100 where the volume bulge is clearly visible.
\label{bulge}}
\end{figure}

Although this is the least technically challenging step of the process, it does present some conceptual difficulties. Our (2+1)-dimensional spacetime has the topology $S^2 \times I$. The timelike interval $I$ runs from $t=1$ to $t=T$ for $T$ time-slices. For computational simplicity the first time slice is identified with the last, changing the topology to $S^2 \times S^1$. When the CDT program is implemented as described in Section~\ref{LQG}, the results are a bulge of extended spacetime and stalks with the minimal number of simplices required to maintain the $S^2$ input topology (Figure \ref{bulge}). The extended bulge performs a random walk along the time axis as we execute sweeps. Since the time axis is a circle, there is no unique choice of time coordinates for the bulge, so it makes little sense to compare a fixed time-slice across sweeps. The solution to this dilemma is to use the maximum volume slice (measured by counting the number of spacelike triangles in the slice) in each sweep as a reference. This ensures that there is at least one common bulk characteristic between any two given slices. See \cite{quant-de-sitter} for a further discussion related to this choice and the issues surrounding it.

\subsection{Step 2: Embed the slices in 3D} \label{P2}

The property of spacetimes generated by the simulation that is of interest is the the intrinsic curvature. This is defined by the deficit angle (\ref{defang}), which is easily measured for individual spacetimes. But because the number and location of vertices will change across sweeps, it is unclear how to measure an expectation value using this quantity. To overcome this problem it is useful to embed the individual spacetimes in three dimensions. The resulting coordinate system can then be used to compare measurements across sweeps. Of course once the spacetime is embedded, it is extrinsic not intrinsic curvature that is being measured. However, Gauss's {\it Theorema Egregium} states that the Gaussian curvature of a two-dimensional surface embedded in a flat three-manifold is intrinsic to the surface, and is completely determined by the intrinsic curvature \cite{gaussTE}. So in 2+1 dimensions, measuring the extrinsic curvature is equivalent to measuring the intrinsic curvature.

\subsubsection{The embedding algorithm} \label{algorithm}.

The points that define vertices in CDT are by necessity coordinate-free. The only information that the simulation generates is the linking between them. In order to embed a two-dimensional spacelike slice in three dimensions for analysis, we have to choose a coordinate for each point based on its relationship to its neighbors. This is accomplished by feeding the logical links between the vertices into a force-directed graphing algorithm, specifically the spring electrical embedding algorithm \cite{graph} as implemented by Mathematica 6.

In this graphing algorithm, neighboring vertices are attracted to each other by a force that is proportional to the Euclidean distance between them, i.e., a spring force. Additionally, every vertex repels every other vertex with a force that falls off with the inverse square of the distance, i.e., an electrical force. The total energy of the system is given by:

\begin{align}								\label{graph-energy}
			E = \dst\sum_{i=1}^{V} \biggl( & -C_{e} \dst\sum_{j \neq i} \frac{x_j - x_i}{d_{ij}^2} \nonumber \\
			& + C_{s} \dst\sum_{\substack{ \mathrm{neighbors} \\ \mathrm{of\ } i}} d_{ij} (x_j - x_i) \biggr)^2 ,
\end{align}		

\noindent where $C_{e}$ and $C_{s}$ are constants that control the relative strength of the electrical and spring forces, respectively, and $d_{ij}$ is the Euclidean distance between vertex $i$ and $j$. This energy function is then minimized by iteratively moving each vertex in the direction of the spring force. The configuration of points that minimizes (\ref{graph-energy}) is used as the final embedding of our spacelike slice.

\subsection{Step 3: Measure distances} \label{P3}

Given an embedding of our surface in flat three-dimensional space, we can, in principle, determine the extrinsic curvature from the positions $\bold r (t)$ of the points. In particular, for a spherical surface, $\bold r (t)$ will be constant, and deviations from a constant value will give a measure of the variation of the curvature. Because of this, a measurement of  $\bold r (t)$ for each of the spacetimes generated by the simulation is therefore a measurement of curvature. Because of this I will focus on the measurement of $\bold r (t)$ for the remainder of this paper.

To perform this measurement, rays emanating from the center of geometry of a particular slice are traced and the point at which they intersect with the manifold is calculated. The rays used are not arbitrary, but rather picked so they will point to specific locations on a 2-sphere which correspond to different pixels in the HEALPix discretization scheme (more on HEALPix in Sec. \ref{P4}). So what is actually measured is the distance from the center of geometry to the two-dimensional surface as a function of HEALPix pixel number or $\bold r (n_{pix})$. In order to avoid correlations that are an artifact of the embedding algorithm, a random rotation is performed about the center of geometry for each set of distance measurements. 

\subsection{Step 4: Spherical harmonic analysis} \label{P4}

\begin{figure}[h]
\centerline{\includegraphics{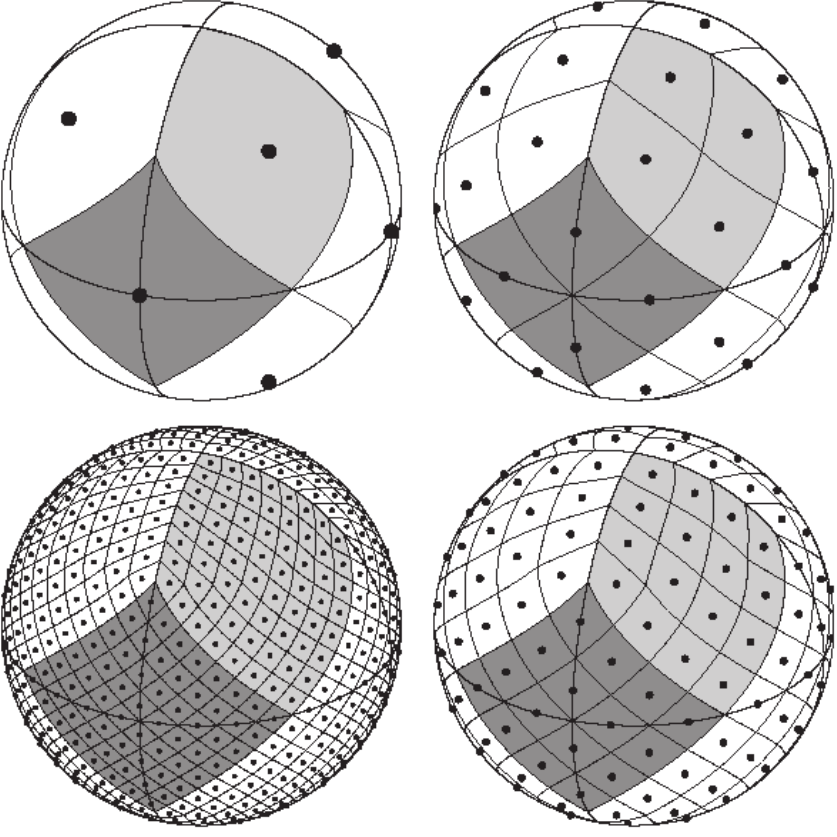}}
\caption{Examples of HEALPix pixelations at different values of $N_{side}$ (image taken from \cite{HEALPix}). The value $N_{side}$ is the number of divisions of the base pixel along one side. Starting at the top left the first sphere is pixelated with the base resolution pixels. The top right sphere is at $N_{side} = 2$. The bottom right sphere is at $N_{side}=4$. The bottom left sphere is at $N_{side}=8$.
\label{HEALPixeg}}
\end{figure}

The measurements in Section~\ref{P3} are taken with a specific method of spherical harmonic analysis in mind. HEALPix (Hierarchical Equal Area iso-Latitude Pixelisation) is an efficient method of pixelating a 2-sphere that lends itself to spherical harmonic transformations. It is used widely in cosmic microwave background analysis, including data from the WMAP and Planck experiment. Details of the HEALPix implementation can be found in \cite{HEALPix, HEALPix2}. The basic idea is that any function $f$ on a sphere can be decomposed into a linear combination of spherical harmonics $Y_{lm}$:

\beq										\label{sh}
			f(\theta,\phi) = \dst\sum_{l=0}^{l_{max}} \dst\sum_m a_{lm} Y_{lm}(\theta,\phi).
\eeq

The $a_{lm}$ can then be found by using the orthogonality of the $Y_{lm}$ to invert (\ref{sh}):

\beq										\label{alm}
			a_{lm}= \int_{\Omega} d\Omega f(\theta,\phi) Y^*_{lm}(\theta,\phi).
\eeq

In the case of a discrete function $\hat f$, evaluated at pixel locations $(\theta_p,\phi_p)$, with $N_{pix}$ pixels, (\ref{alm}) can be approximated by a sum. The approximation used by HEALPix is \cite{HEALPix2}

\beq										\label{almpix}
			\hat a_{lm}= \frac{4 \pi}{N_{pix}} \dst\sum_{p=0}^{N_{pix}-1} Y^*_{lm}(\theta_p,\phi_p)\hat f (\theta_p,\phi_p).
\eeq

The variance of the $\hat a_{lm}$ is the angular power spectrum, $\hat C_l$:

\beq										\label{eq:cl}
			\hat C_l = \frac{1}{2 l +1} \dst\sum_{m} \vert \hat a_{lm} \vert^2.
\eeq

The resolution of HEALPix maps is characterized by the quantity $N_{side}$ which represents the number of divisions of a base-resolution pixel \cite{HEALPix}. Examples of pixelations with different values of $N_{side}$ are shown in Figure \ref{HEALPixeg}.

\subsection{Errors in the embedding} \label{P2-errors}

Although the embedding algorithm does a very good job, it is not perfect. The requirement of Section \ref{CDT} that all link lengths be equal is very difficult for the algorithm to achieve; most links are the same length, but there are always some that are either bigger or smaller than the average. When the link lengths are different sizes, the geometry of our spacelike slices becomes distorted. This means that there is some difference between the actual geometry of our manifold and the geometry we measure in step 3 of our process. To understand how this will effect our measurements, we look at how each triangle that is intersected by a ray is distorted and apply an extra amount $\Delta r$ to our measurements. We estimate $\Delta r$ as follows: the change in the area of a triangle with changes in link length $l$ is roughly $\Delta A \sim l \Delta l$. The change in the area of a spherical spacetime with changes in the radius $r$ is roughly $A \sim r \Delta r$. Since the spherical spacetimes are constructed of triangles and the radius $r$ roughly corresponds to our ray-trace measurements, we can equate these two quantities to get an estimate of $\Delta r$: $\Delta r \sim \frac{l}{r} \Delta l$, where $l$ is the average link length for a given spacetime, $\Delta l$ is is the difference of the average of a specific triangle's link lengths and the average link length for the spacetime, and $r$ is the measured distance to the triangle in question.

This quanity $\Delta r$ is calculated for each pixel and applied to the original measurement. Then the new distances are are analyzed using the methods described in Section \ref{P4}. The results of this additional analysis are included in Figures \ref{fig:powers} and \ref{fig:fluctpowers}, which are discussed in Section \ref{results}.

\section {Results} \label{results}

\begin{figure}[h]
\centerline{\includegraphics{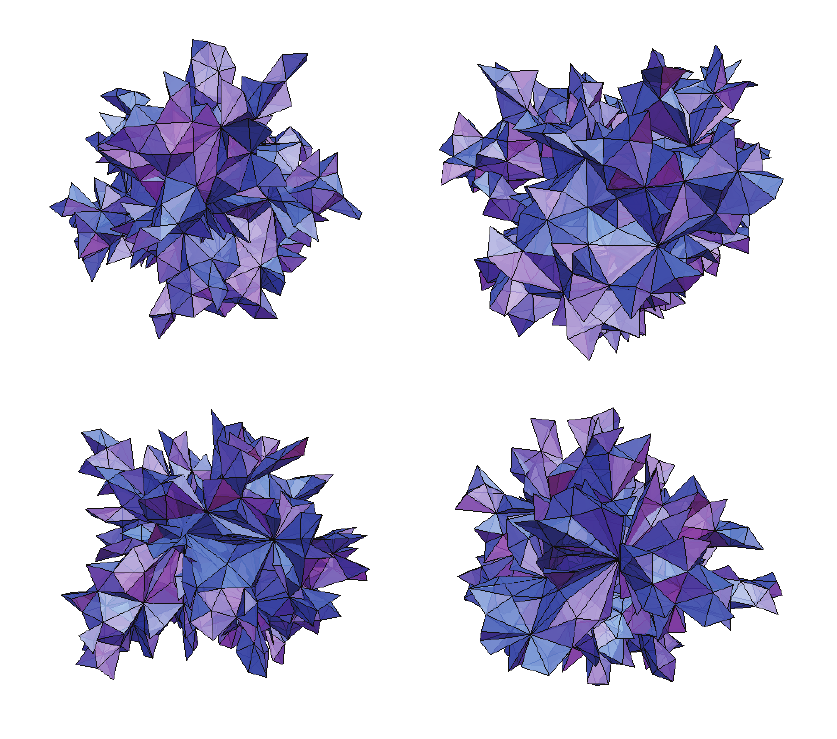}}
\caption{Various max volume spacelike slices from simulation 1 embedded using the algorithm described in section \ref{P2}. Top left is sweep 55500, top right is sweep 71800, bottom left is sweep 92700 and bottom right sweep 99900.
\label{fig:sweepembeds}}
\end{figure}

In order to test the sphericity of the output from CDT, five simulations were run. The values of the bare coupling constants for each of these simulations are listed in Table \ref{ex:simvals}. A ``snapshot" of the spacetime was exported to a file every 100 sweeps. The largest volume spacelike slice was then chosen from each of these spacetimes and used for analysis. Examples of several of these slices, embedded in three-dimensions, are shown in Figure \ref{fig:sweepembeds}.

\begin{table*}[!]
\begin{center}

\begin{tabular}{|c|ccr@{.}lr@{.}lr@{.}lccp{2.0in}r@{.}l|}
	\hline
	\textbf{Sim}\
	& $V_{init}$
	& $T$
	& \multicolumn{2}{c}{$k_0$}
	& \multicolumn{2}{c}{$k_3$}
	& \multicolumn{2}{c}{$\epsilon$}
	& Total sweeps
	& Themalized sweep
	& Total samples(\# of subsamples)
	& \multicolumn{2}{c}{$\delta S^2$} \vline
	\\ \hline \hline
	1	&	81921	&	64	&	1&0	&	0&78	&	0&02	&	100000	&	46300	&	538	&	0&055 \\
	2	&	81921	&	64	&	1&5	&	0&86	&	0&02	&	100000	&	70100	&	300	&	0&06\\
	3	&  	81921	&	64	&	2&0	&	0&94	&	0&02	&	400100	&	24800	&	3745; 1873(50); 1249(50); 937(50); 749(50); 625(50)	&	0&061 \\
	4	&  	81921	&	64	&	2&5	&	1&03	&	0&02	&	100000	&	17800	&	822	&	0&069	\\
	5	&  	81921	&	64	&	3&0	&	1&12	&	0&02	&	100000	&	19400	&	807	&	0&06 \\
	\hline
	
\end{tabular}
\end{center}
\caption{Various parameters and results for each simulation. $V_{init}$ is the total initialization volume (i.e. the number of initial simplices), $T$ is the number of time-slices, $k_0$ and $k_3$ are the bare coupling constants discussed in Section \ref{CDT}. The numbers in parentheses next to the total samples for simulation 3 are the number of random subsamples of the full set of sweeps that were made. These results were then averaged to avoid sample bias. $\delta S^2$ is the deviation from being perfectly spherical discussed in Section \ref{spacetime-shape}.}
\label{ex:simvals}
\end{table*}

\begin{figure}[h]
\centerline{\includegraphics[width=0.5\textwidth]{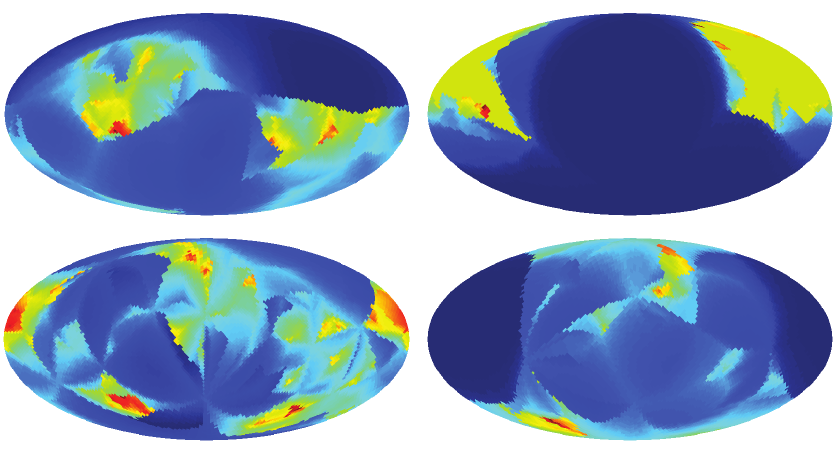}}
\caption{The HEALPix maps for the slices in Figure \ref{fig:sweepembeds}. Top left is sweep 55500, top right is sweep 71800, bottom left is sweep 92700 and bottom right sweep 99900.
\label{fig:sweepmaps}}
\end{figure}

\begin{figure}[h]
\centering
\subfloat{\includegraphics[width=0.25\textwidth]{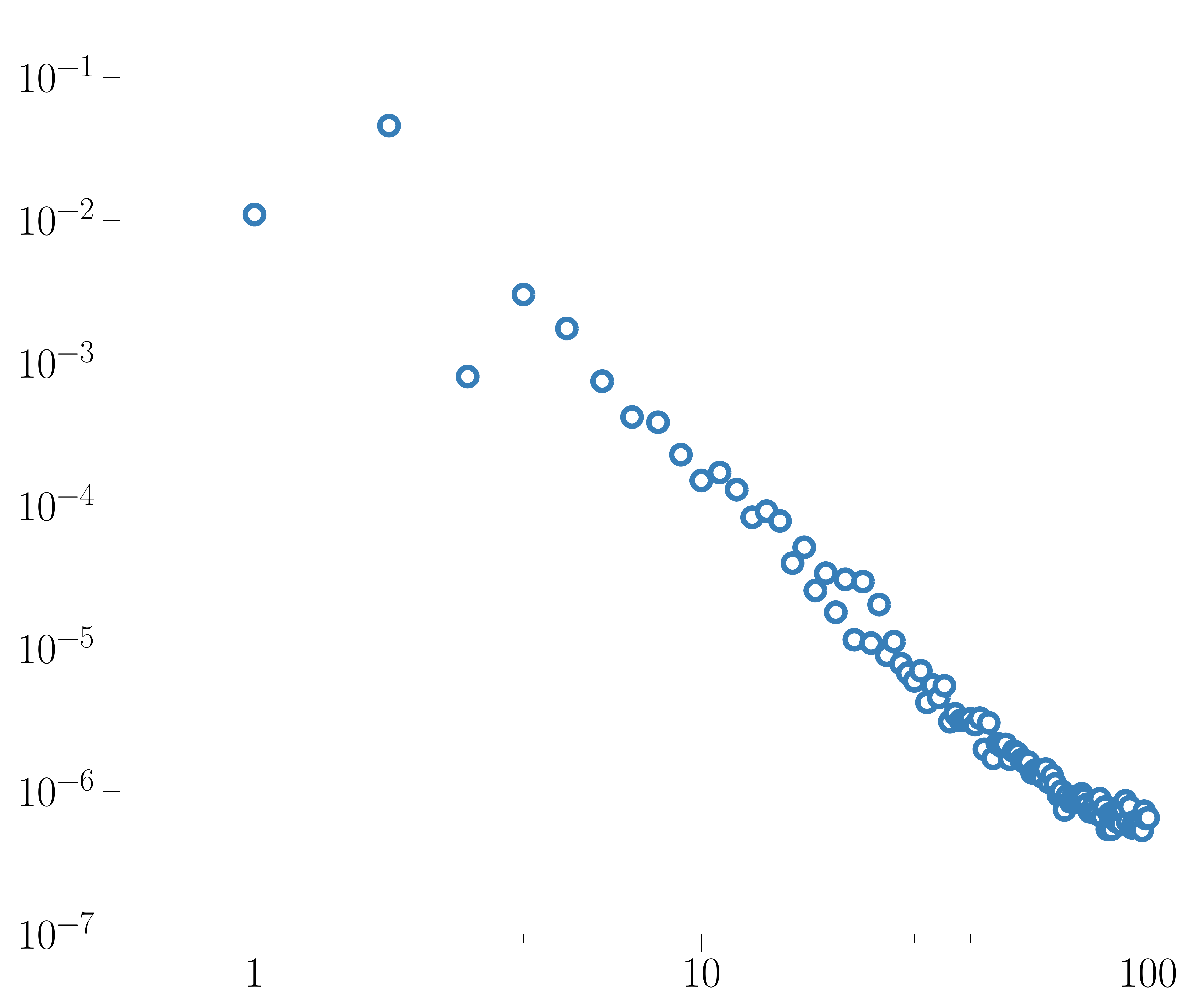}}
\subfloat{\includegraphics[width=0.25\textwidth]{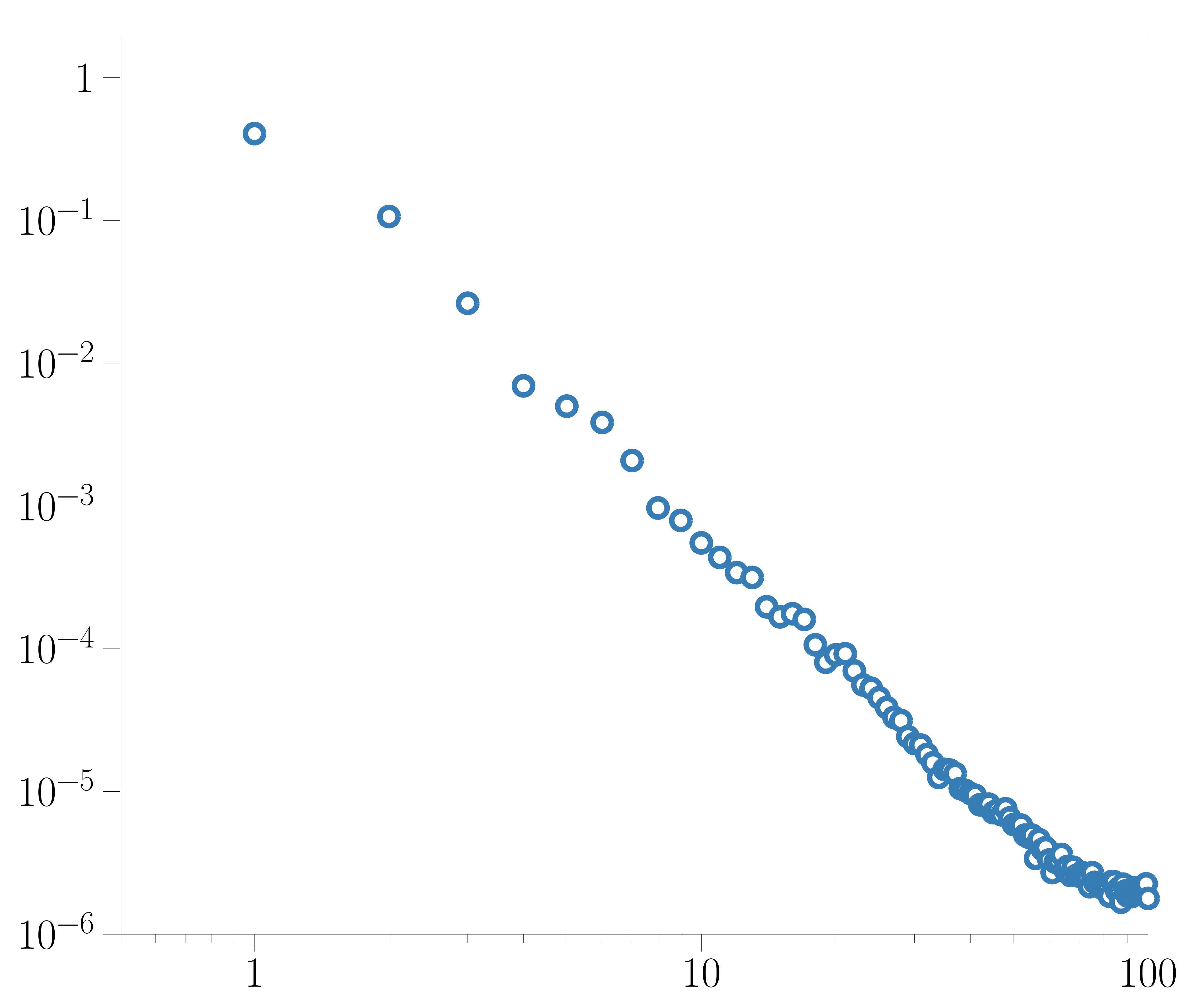}} \\
\subfloat{\includegraphics[width=0.25\textwidth]{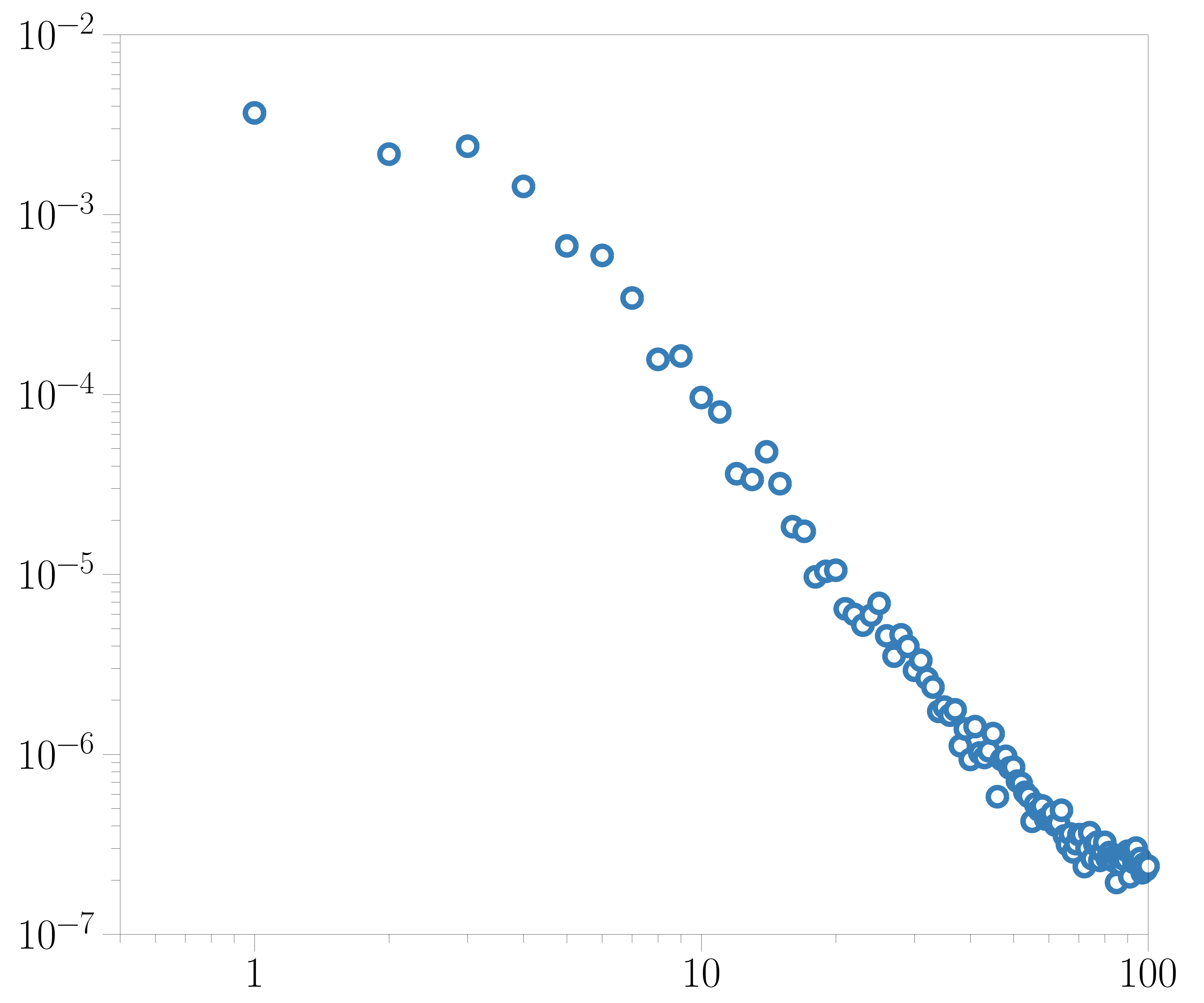}} 
\subfloat{\includegraphics[width=0.25\textwidth]{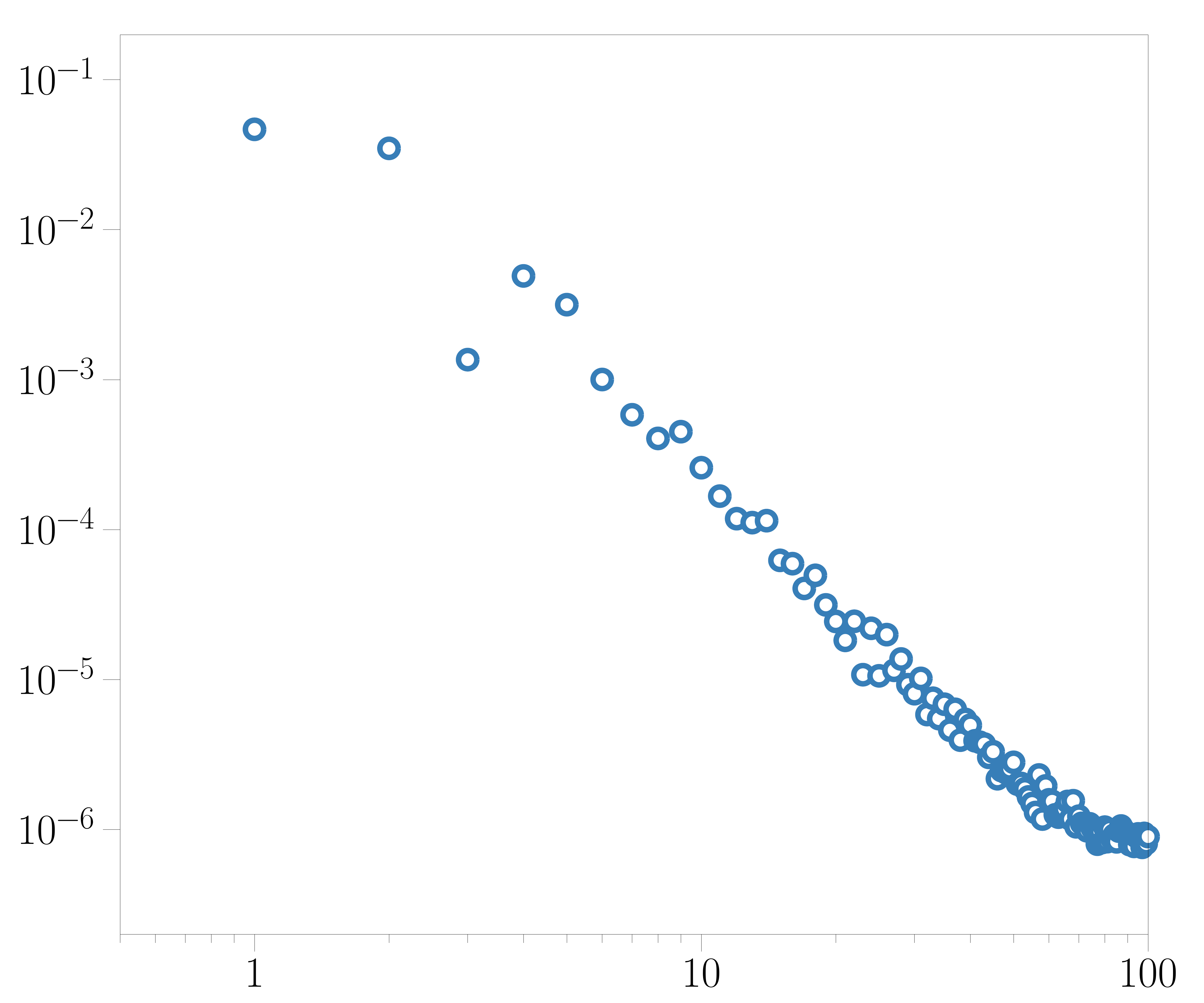}} 
\caption{$C_l/C_0$ vs. $l$ for each of the individual spacetimes in Figure \ref{fig:sweepembeds}. Top left is sweep 55500, top right is sweep 71800, bottom left is sweep 92700 and bottom right sweep 99900.
\label{fig:sweeppowers}}
\end{figure}

All of the HEALPix analysis was performed at $N_{side}=$ 32. This resolution effectively limits the power spectrum analysis to the first 100 $l$'s. At higher values than this, we start sampling the map at scales smaller than the map resolution, which results in noise in the spectrum.  The power spectrum, up to $l=100$, and HEALPix map for each of the slices in Figure~\ref{fig:sweepembeds}, are shown in Figure~\ref{fig:sweepmaps} and Figure~\ref{fig:sweeppowers} respectively. 

\subsection{The shape of spacetime} \label{spacetime-shape}

As can be seen in Figure \ref{fig:sweeppowers}, individual slices vary from being spherical by non-trivial amounts. However, individual slices come from particular paths in the path integral (\ref{pic}), and physically, only the ensemble averages are meaningful. It is important to choose carefully when picking the quantities from which to calculate these averages. Because all notion of absolute location is washed out from sweep to sweep, averaging the distance measurements is an unreliable measure of the shape of the ensemble system. Each pixel will explore roughly the same range of allowed values, resulting in the same average distance everywhere -- the measurements will ``sum to zero.'' If there were a non-spherical shape preferred by the system, it would be washed out in this average. We need a quantity that is rotationally invariant, i.e., some measurement that will give us the same answer for a given shape regardless of how that shape is oriented relative to a coordinate system. Luckily, the power spectrum is just such a quantity. Averaging each $l$ of the power spectrum across the entire ensemble will give us a measure of the physical shape of our spacetime.

To test this assumption, simulation 3 was run for four times as many sweeps as the other simulations. This pool of sweeps was then sampled at different sizes. As sample size increases, measurements that are subject to the ``sum to zero'' effect will converge to average values everywhere. Measurements that are free from this defect will remain unaffected. If the power spectrum ensemble average is one of these quantities, then it should be almost the same at any reasonably large sample size. Figure \ref{fig:sampled_cl} confirms this. The averaged power spectra are almost identical for every $l$ at all sample sizes.

\begin{figure*}[h]
\centering
\subfloat{\includegraphics[width=0.5\textwidth]{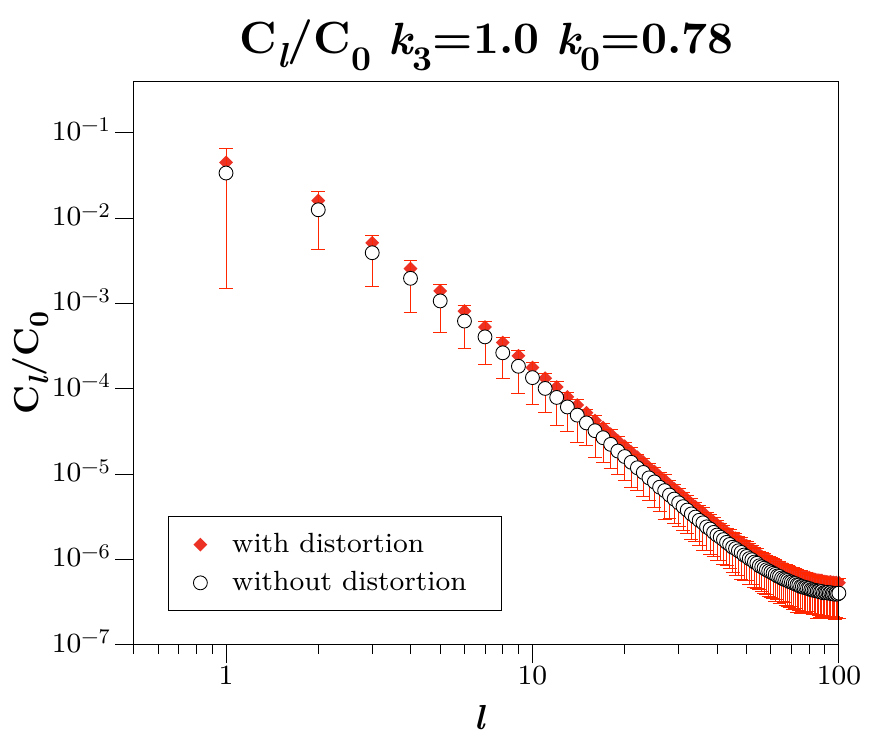}}
\subfloat{\includegraphics[width=0.5\textwidth]{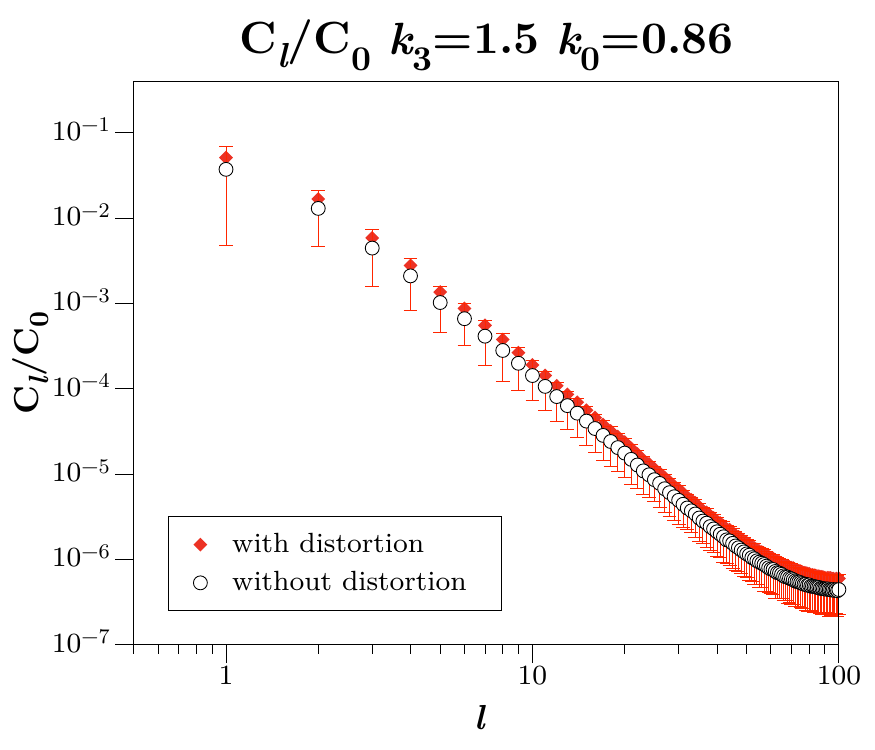}} \\
\subfloat{\includegraphics[width=0.5\textwidth]{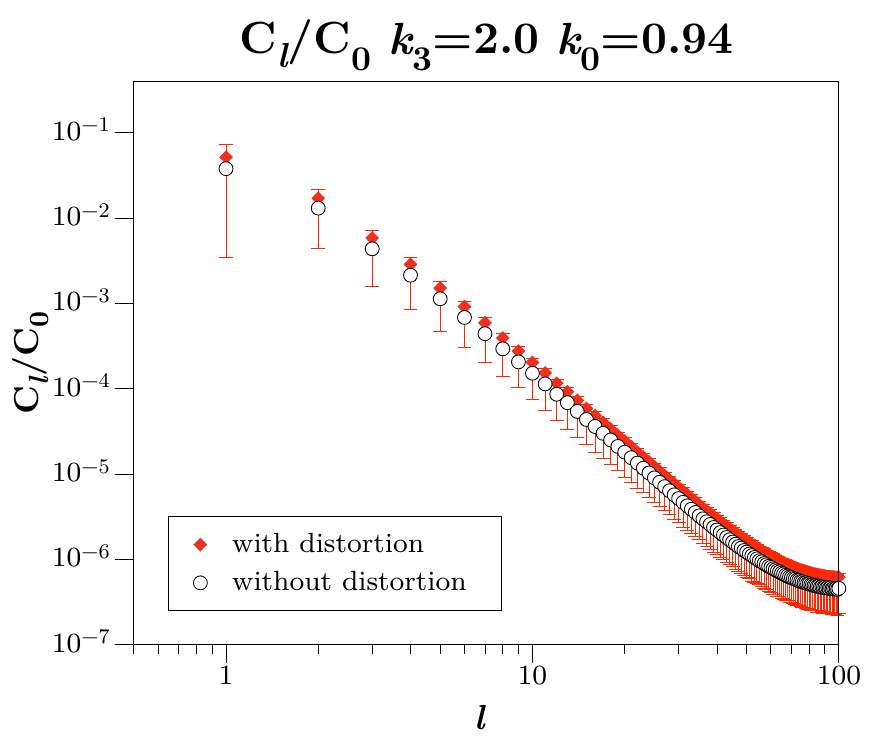}} 
\subfloat{\includegraphics[width=0.5\textwidth]{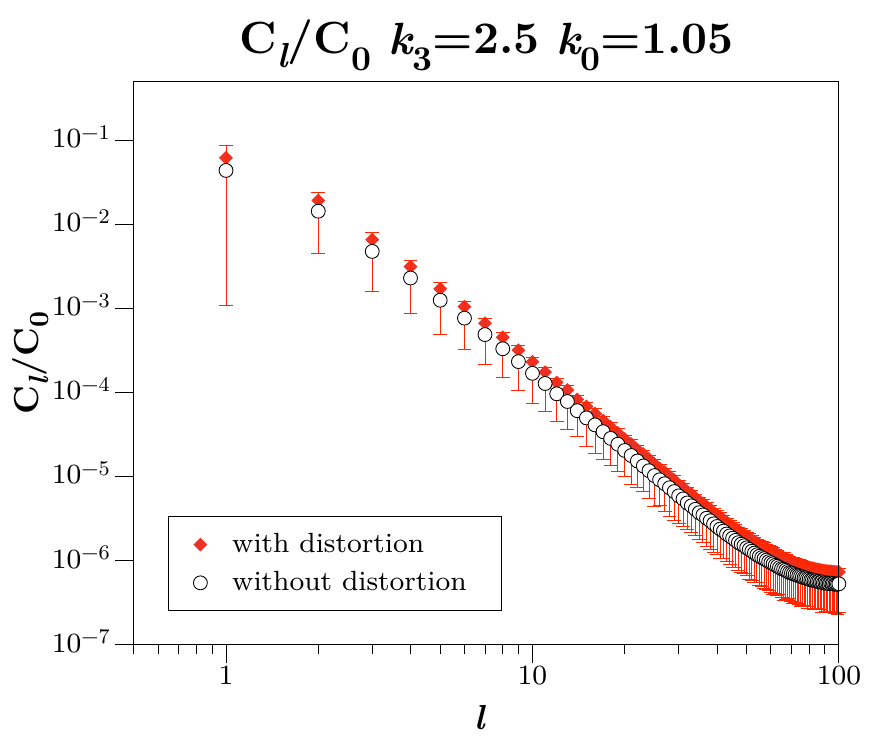}} \\
\subfloat{\includegraphics[width=0.5\textwidth]{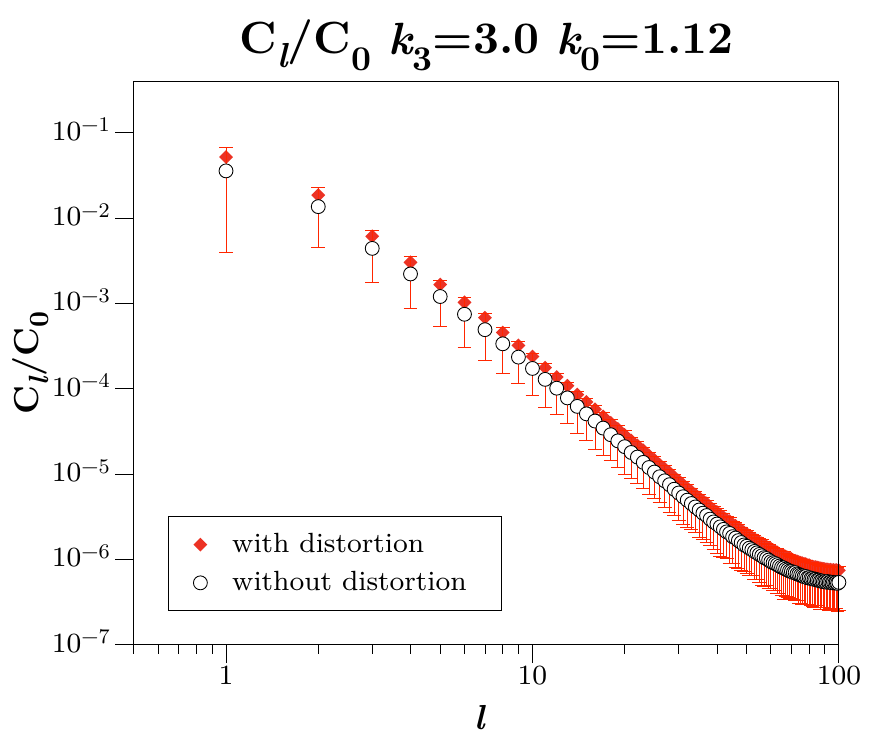}}
\caption{The relative power of modes greater than $l=0$ when testing the ensemble shape of the spacetime for each of the simulations for measurements with and without the distortion described in Section \ref{P2-errors} applied. As can be seen from the plots the distortion induced by embedding errors is small.
\label{fig:powers}}
\end{figure*}

Now that we have some confidence in our averaging procedure, we can examine the results for all of the simulations in Table~\ref{ex:simvals}. The results are shown in Figure~\ref{fig:powers}. All of the simulations performed here seem to be very spherical. In each case, the $l=1$ mode is only a few percent of the the spherical or $l=0$ mode, and higher modes fall off rapidly. Also, the results are very similar for all chosen values of $k_0$ and $k_3$, implying that the shape of the spacetime  in the extended phase does not depend on the values of the coupling constants. Finally, in order to give each simulation a single score to rank its sphericity, the following quantity is calculated:

\beq										\label{sphericity}
			\delta S^2= \frac { \dst\sum_{l=1}^{100} C_l} { C_0 }.
\eeq

$\delta S^2$ is a measure of how much power is not in the spherical $l=0$ mode. The results for each spacetime are shown in Table \ref{ex:simvals}.

\subsection{The fluctuations} \label{fluctuations}

The spacetimes produced by CDT seem quite spherical, matching the results from canonical quantization. The other prediction of the canonical procedure is that the spacetime should have no fluctuations. To test whether the CDT spacetimes exhibit this property I will look at the variance of the distance measurements:

\beq										\label{eq:variance}
			\sigma_{pix}^2 = \langle r(n_{pix})^2\rangle-\langle r(n_{pix})\rangle^2.
\eeq

\begin{figure*}[h]
\centering
\subfloat[Shape power spectra at different sample sizes]{\label{fig:sampled_cl}\includegraphics[width=\textwidth]{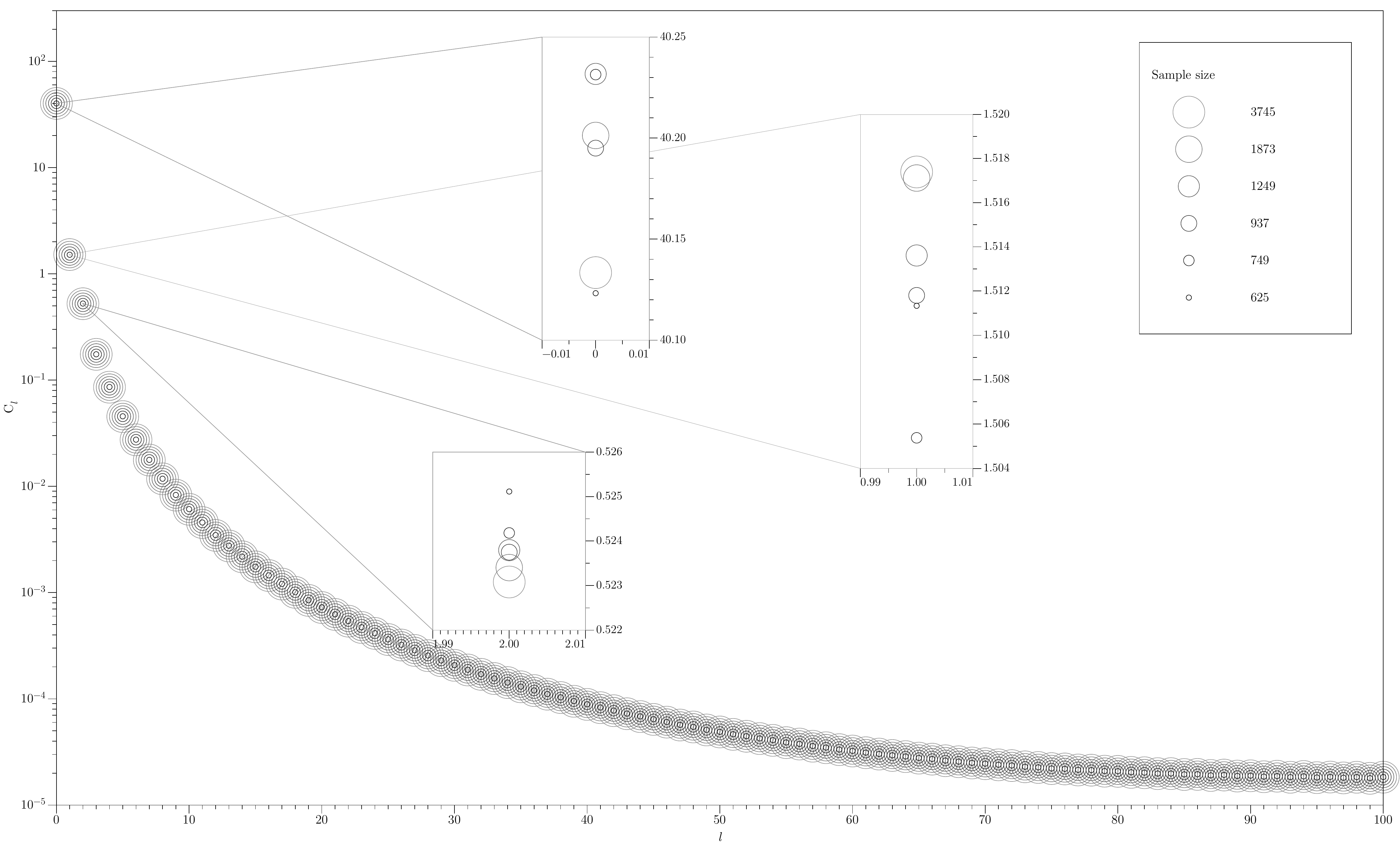}} \\
\subfloat[Shape fluctuation power spectra at different sample sizes]{\label{fig:sampled_distvar}\includegraphics[width=\textwidth]{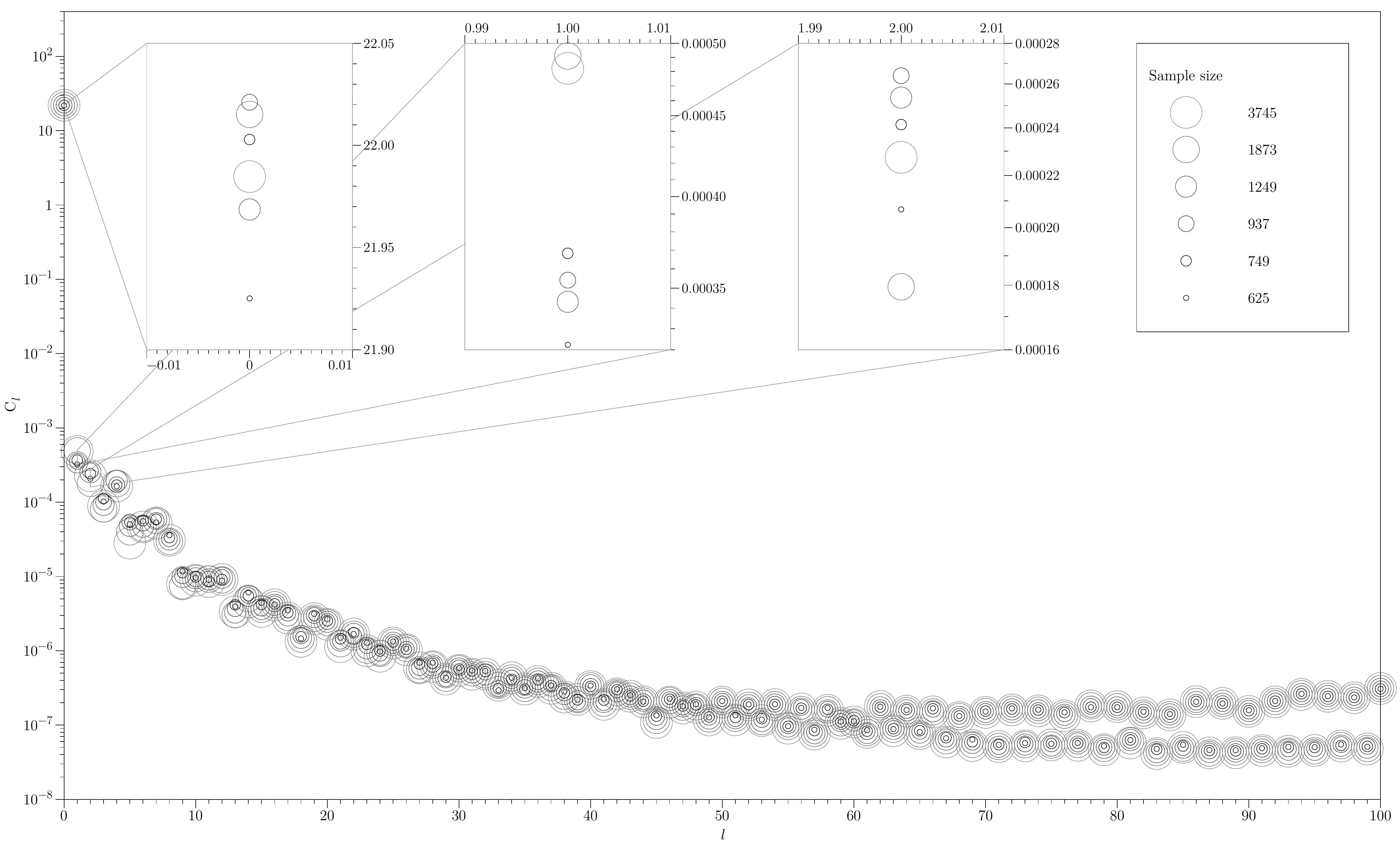}} \\
\caption{The effects of sample size on the measurement of the shape and the shape fluctuations.
\label{fig:samples}}
\end{figure*}

\noindent As with the measurement in Section \ref{spacetime-shape}, it is important to note that (\ref{eq:variance}) is not affected by the ``sum to zero'' effect, as can be seen in Figure \ref{fig:sampled_distvar}.

\begin{figure*}[h]
\centering
\subfloat{\includegraphics[width=0.5\textwidth]{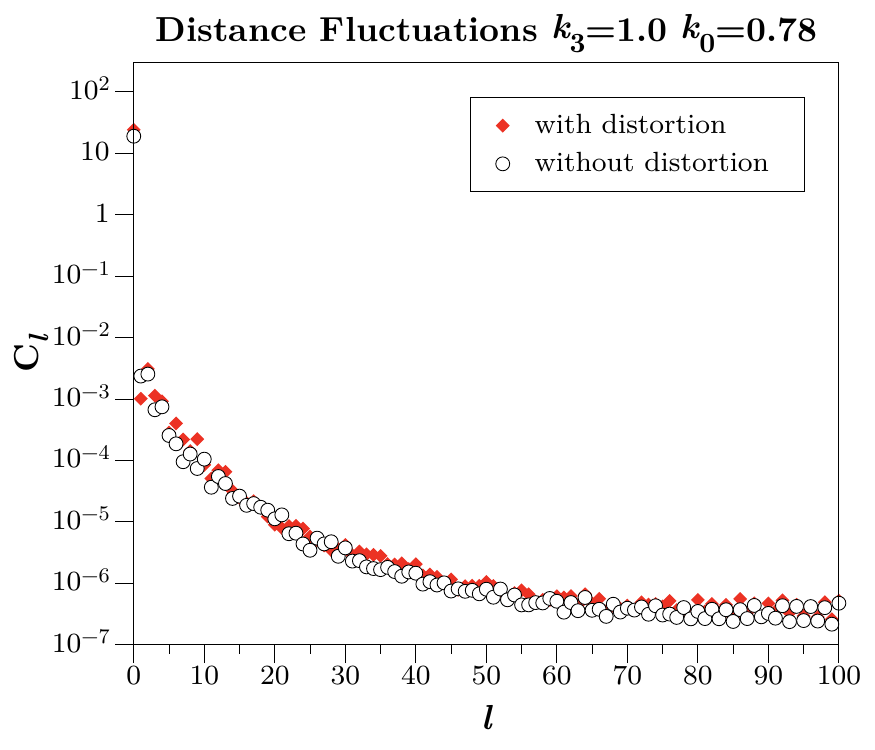}}
\subfloat{\includegraphics[width=0.5\textwidth]{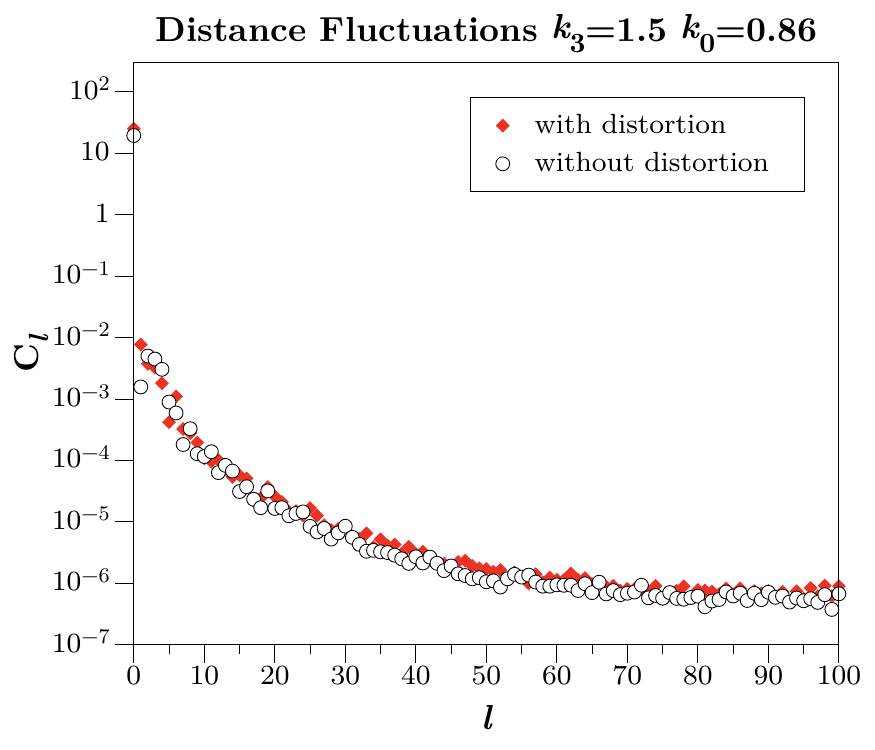}} \\
\subfloat{\includegraphics[width=0.5\textwidth]{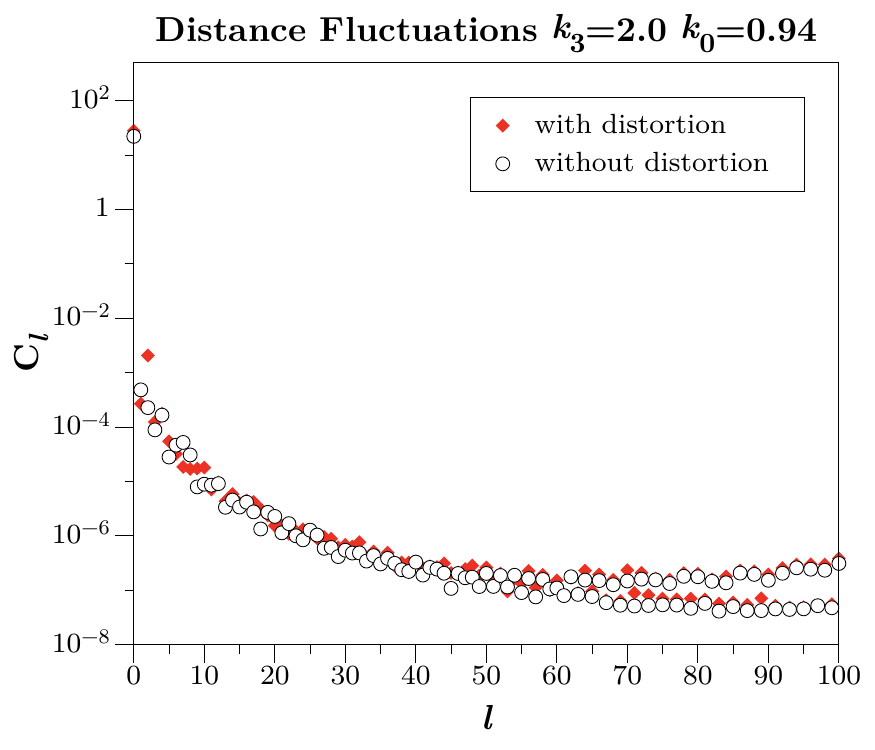}} 
\subfloat{\includegraphics[width=0.5\textwidth]{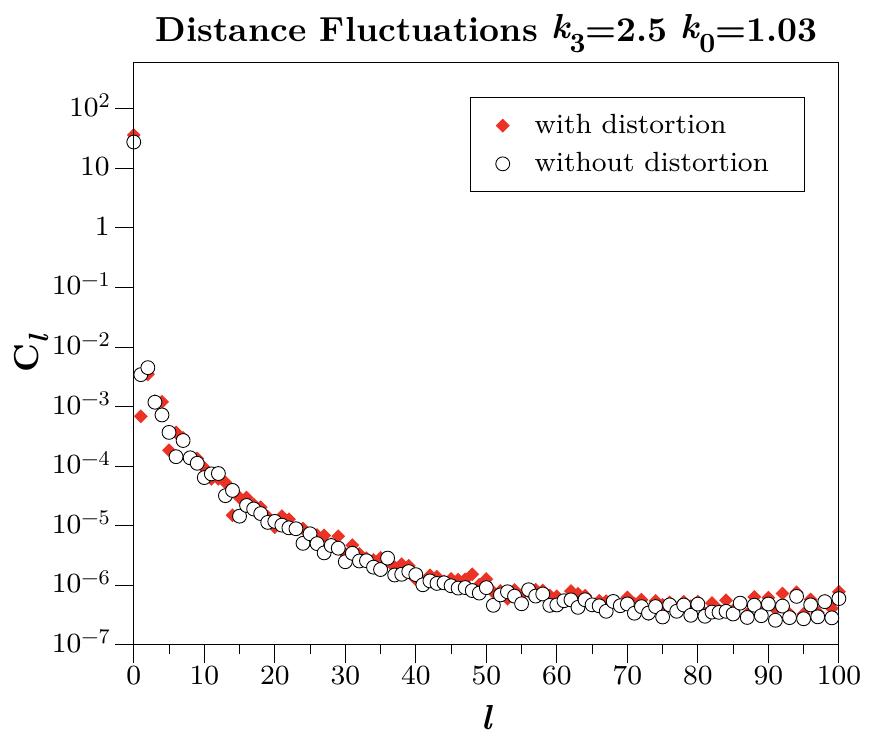}} \\
\subfloat{\includegraphics[width=0.5\textwidth]{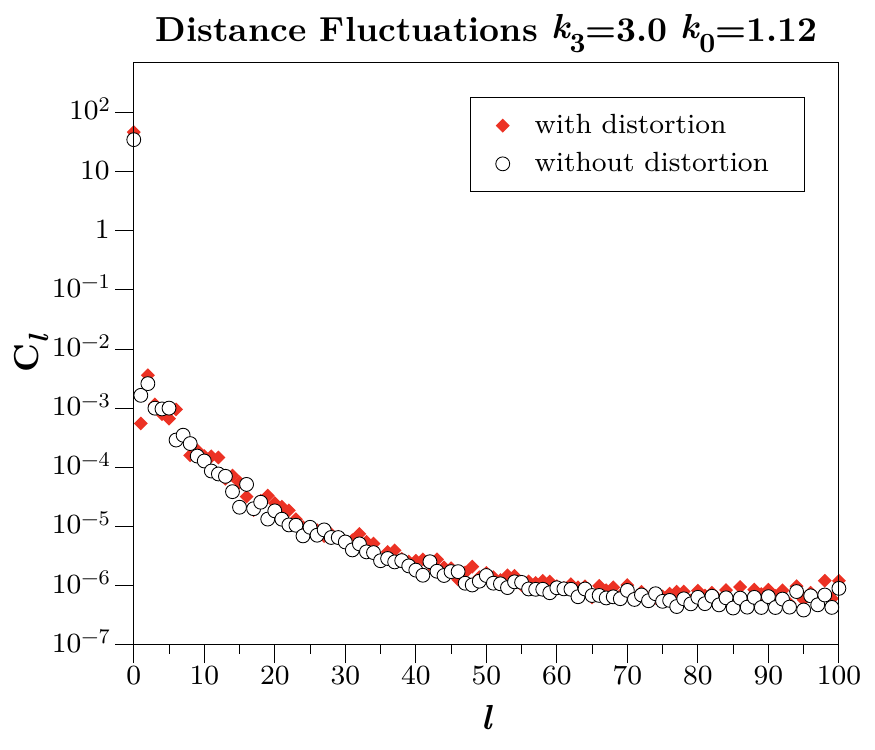}}
\caption{The total power of ensemble spacetime shape fluctuations for each simulation with and without the distortion described in Section \ref{P2-errors} applied. As can be seen from the plots the distortion induced by embedding errors is small. 
\label{fig:fluctpowers}}
\end{figure*}

\begin{figure*}[h]
\centering
\subfloat[$k_0=1.0$]{\includegraphics[width=0.5\textwidth]{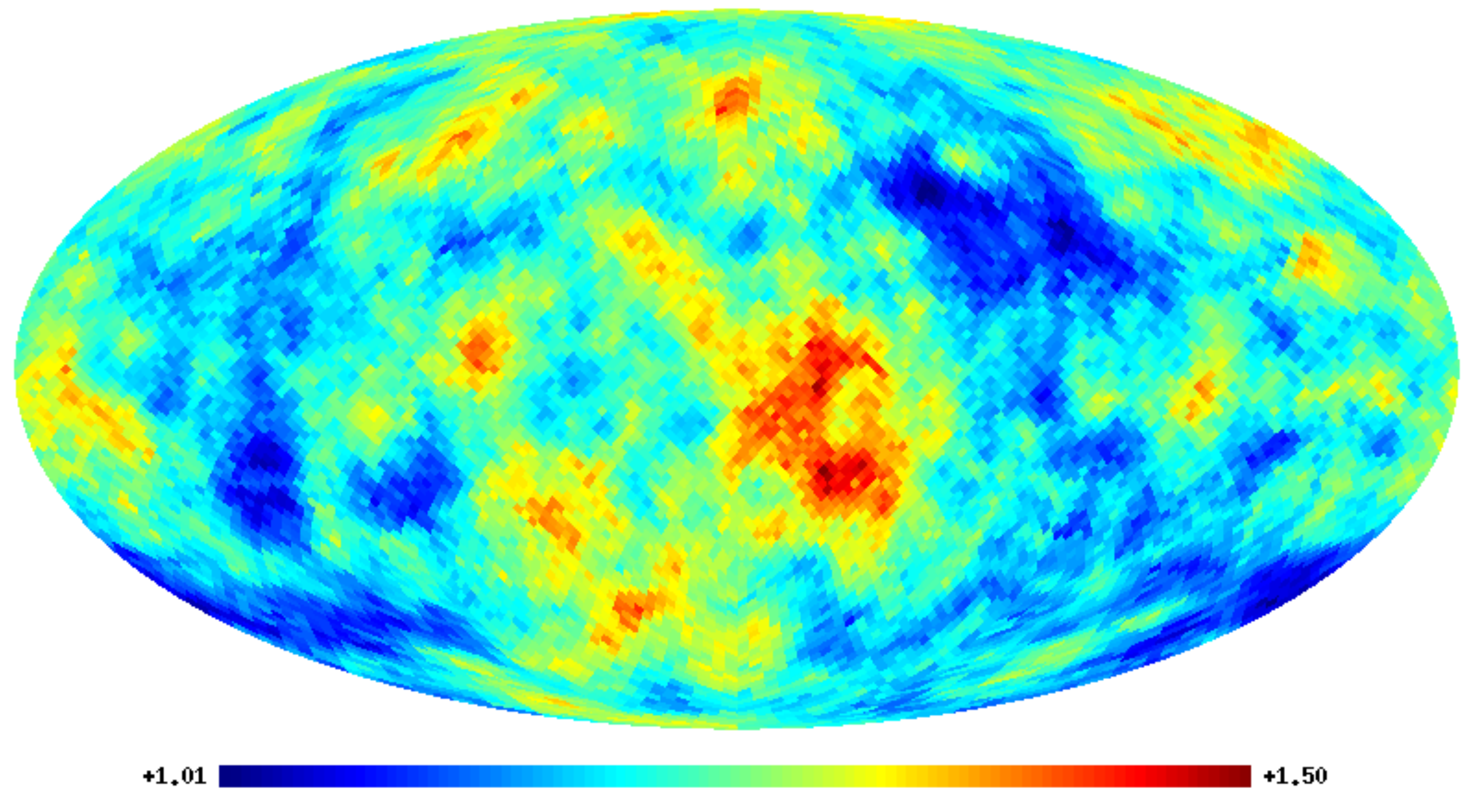}}
\subfloat[$k_0=1.5$]{\includegraphics[width=0.5\textwidth]{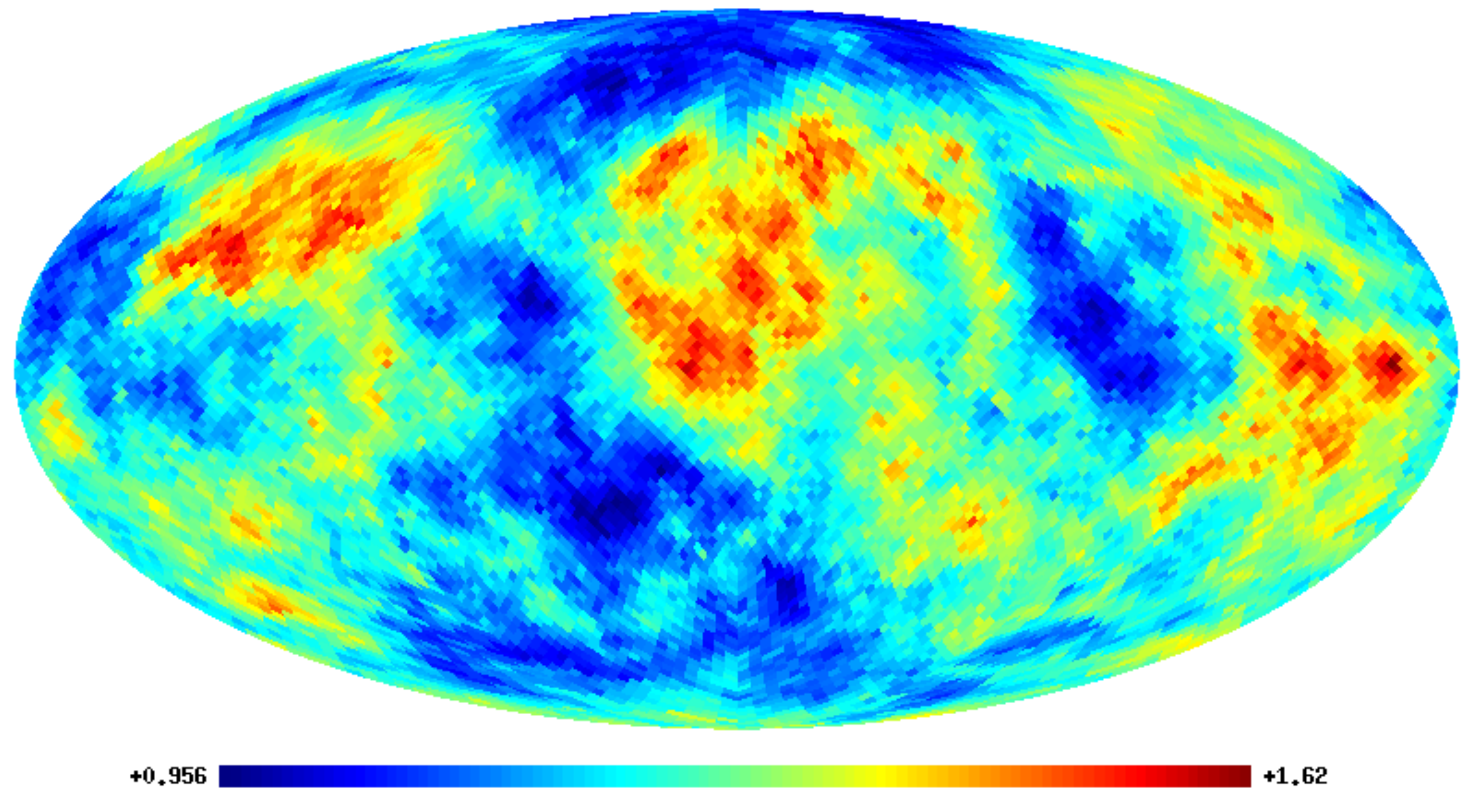}} \\
\subfloat[$k_0=2.0$]{\includegraphics[width=0.5\textwidth]{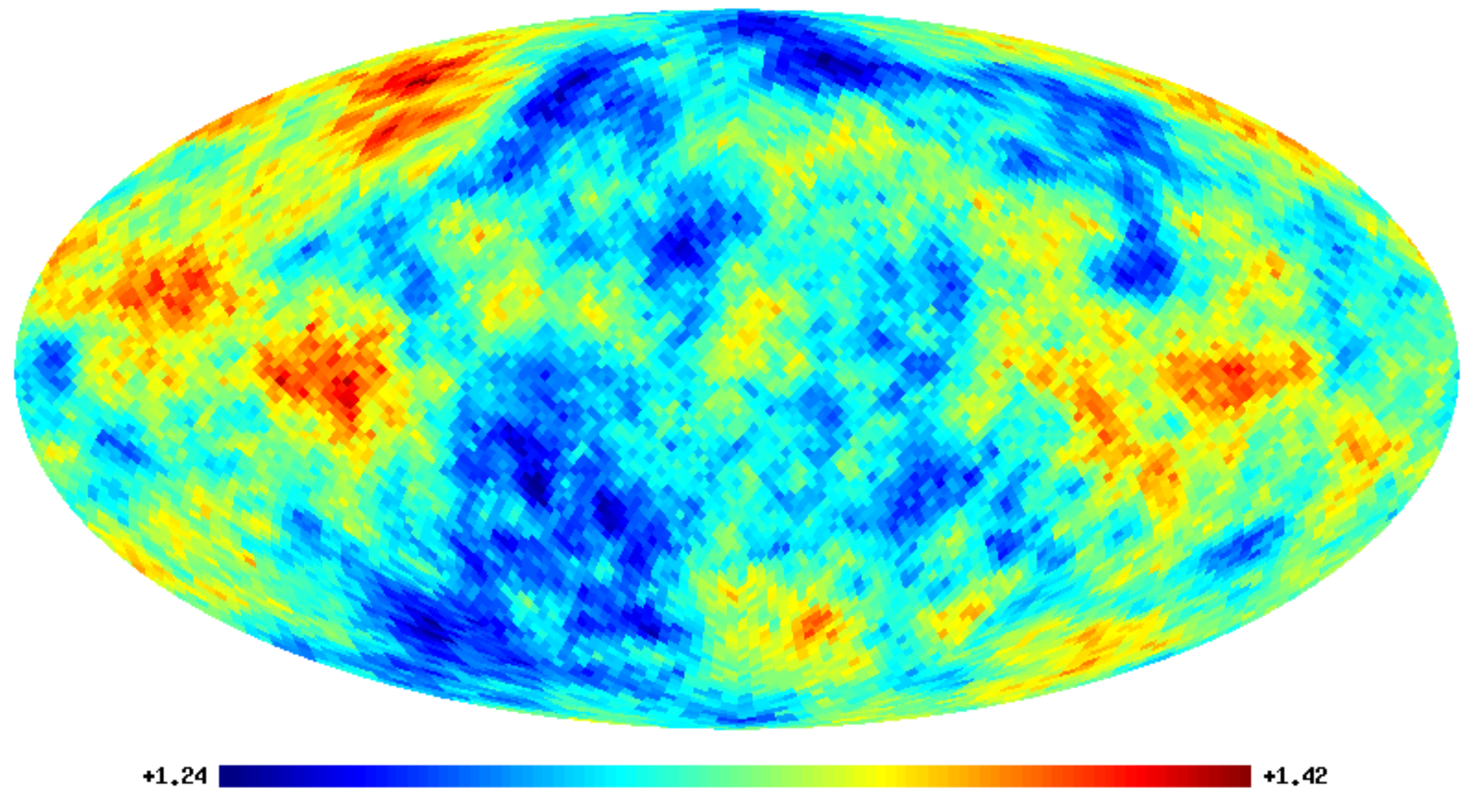}} 
\subfloat[$k_0=2.5$]{\includegraphics[width=0.5\textwidth]{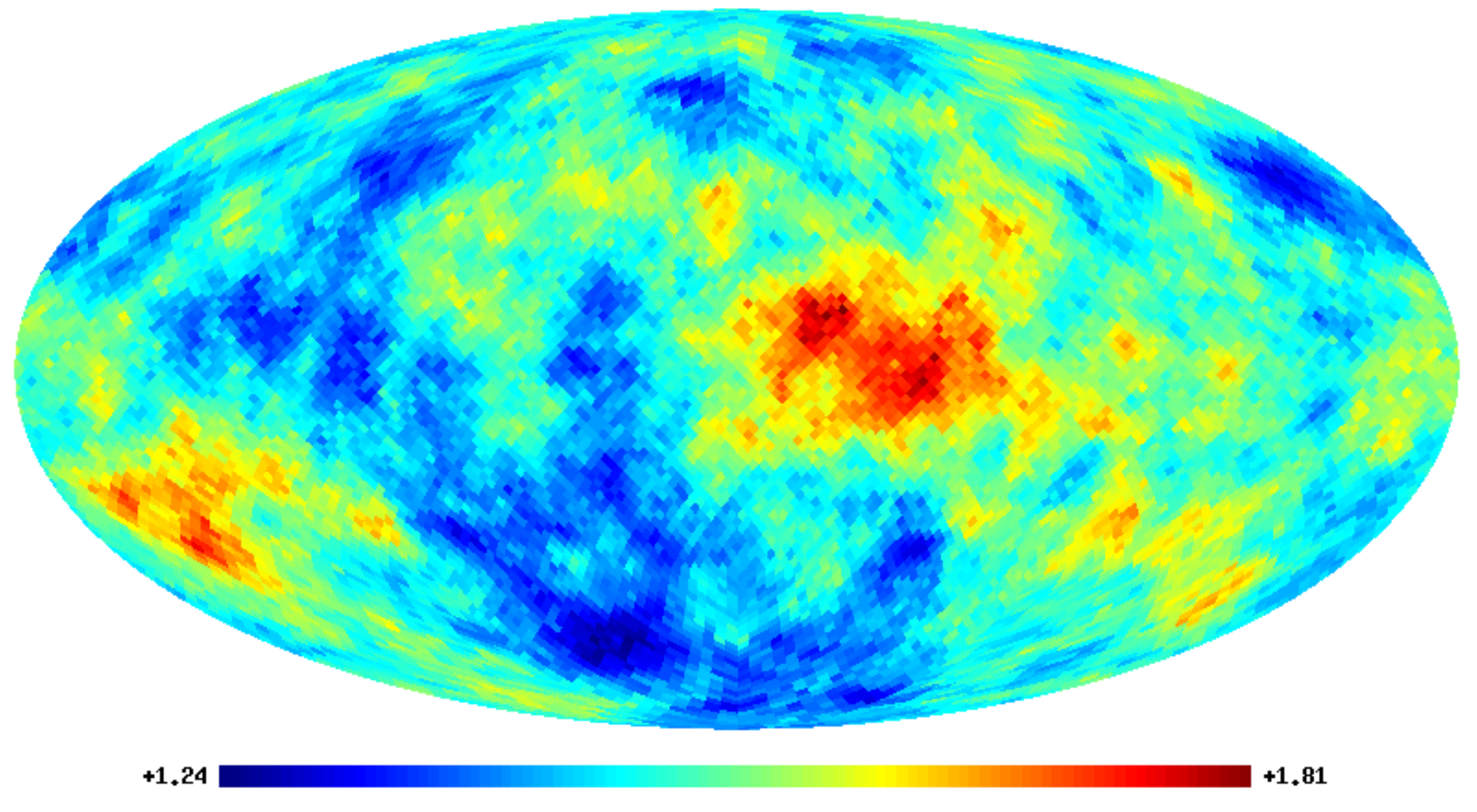}} \\
\subfloat[$k_0=3.0$]{\includegraphics[width=0.5\textwidth]{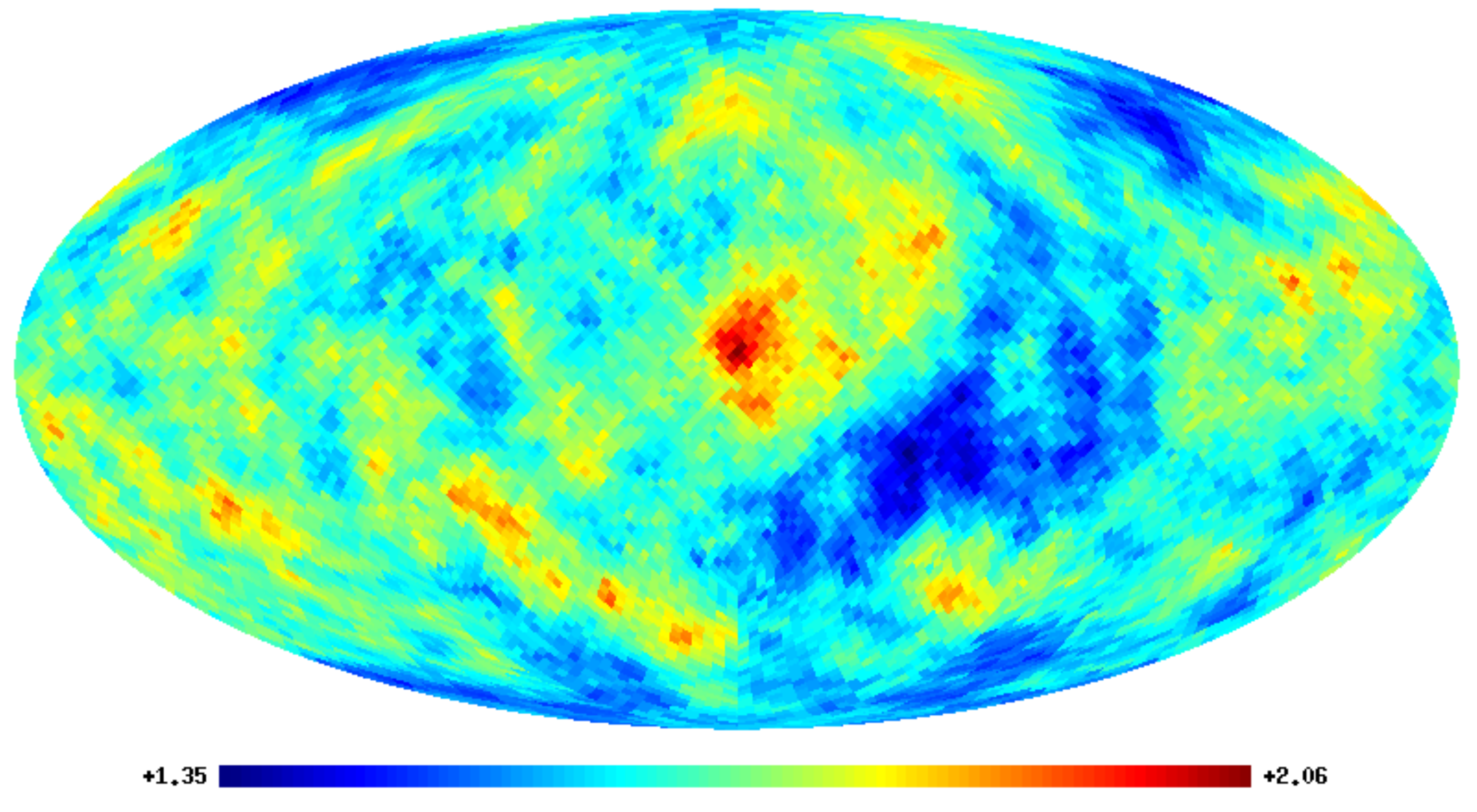}}
\caption{Maps of the ensemble spacetime shape fluctuations for each simulation.
\label{fig:fluctmaps}}
\end{figure*}

Equation (\ref{eq:variance}) is calculated for every pixel in the HEALPix map, creating a map of the fluctuations of the spacetime (Figure \ref{fig:fluctmaps}). A power spectrum is then calculated for this map. The result for each simulation can be seen in Figure \ref{fig:fluctpowers}. Clearly non-spherical (non $l=0$) modes of fluctuations are very small, a result that fits with the predictions of the canonical quantization procedure. There is however, power in the the $l=0$ mode that does not match the canonical predictions. This zero mode power can be attributed to lattice effects in the CDT model. Because the slices are discrete, when the largest volume time-slice is selected (see Section \ref{P1}) it will typically be slightly smaller than the maximum volume slice in the limit where the number of time-slices gets very large. This means that the volume I sample will fluctuate around its ``maximum'' value. Because the maximum volume is fluctuating and our ensemble spacetimes are very spherical, the radius of the spacetime will fluctuate uniformly, resulting in a large zero mode in the fluctuation power spectrum. To calculate an order of magnitude estimate for this fluctuation, the points marking the volumes of the time-slices around the maximum volume time slice were fit to a curve using a cubic curve-fitting algorithm. The maximum of this curve was then interpreted to be the ``actual'' maximum volume in the limit where the number of time-slices becomes large. The average difference between the ``actual'' value and the selected value was calculated. In order to compare this average $\Delta r$ to the zero mode of the power spectrum, one must relate the $C_l$ to the total power in a function $f$ via Parseval's theorem:

\beq										\label{eq:spherepower}
			\frac{1}{4 \pi}  \int_{\Omega} f(\Omega)^2 d\Omega =  \dst\sum_{l=0}^{\infty} \dst\sum_{m=-l}^{l} a_{lm}^2.
\eeq

\noindent Combining this with (\ref{eq:cl}) gives:

\beq										\label{eq:spherepowercl}
			\frac{1}{4 \pi}  \int_{\Omega} f(\Omega)^2 d\Omega =  \dst\sum_{l=0}^{\infty} (2 l + 1) C_l.
\eeq

\noindent The function we are considering is the variance (\ref{eq:variance}), which we will assume is constant over the sphere, and because it is so much bigger, the only $C_l$ that we need to look at is $C_0$, so:

\beq										\label{eq:c0}
			\sigma_r^4 \sim C_0.
\eeq

Additionally, since we are assuming that the spacetime is mostly spherical $r \sim \sqrt{V}$ and $\Delta r \sim \sqrt{V_{max}} - \sqrt{V_{measured}}$ (we are working in two dimensions, so the volume in this case is actually an area). So we have

\beq										\label{eq:c0deltar}
			\Delta r \sim \sqrt{V_{max}} - \sqrt{V_{measured}} \sim C_0^{1/4}.
\eeq

\begin{figure}[h]
\centerline{\includegraphics[width=0.5\textwidth]{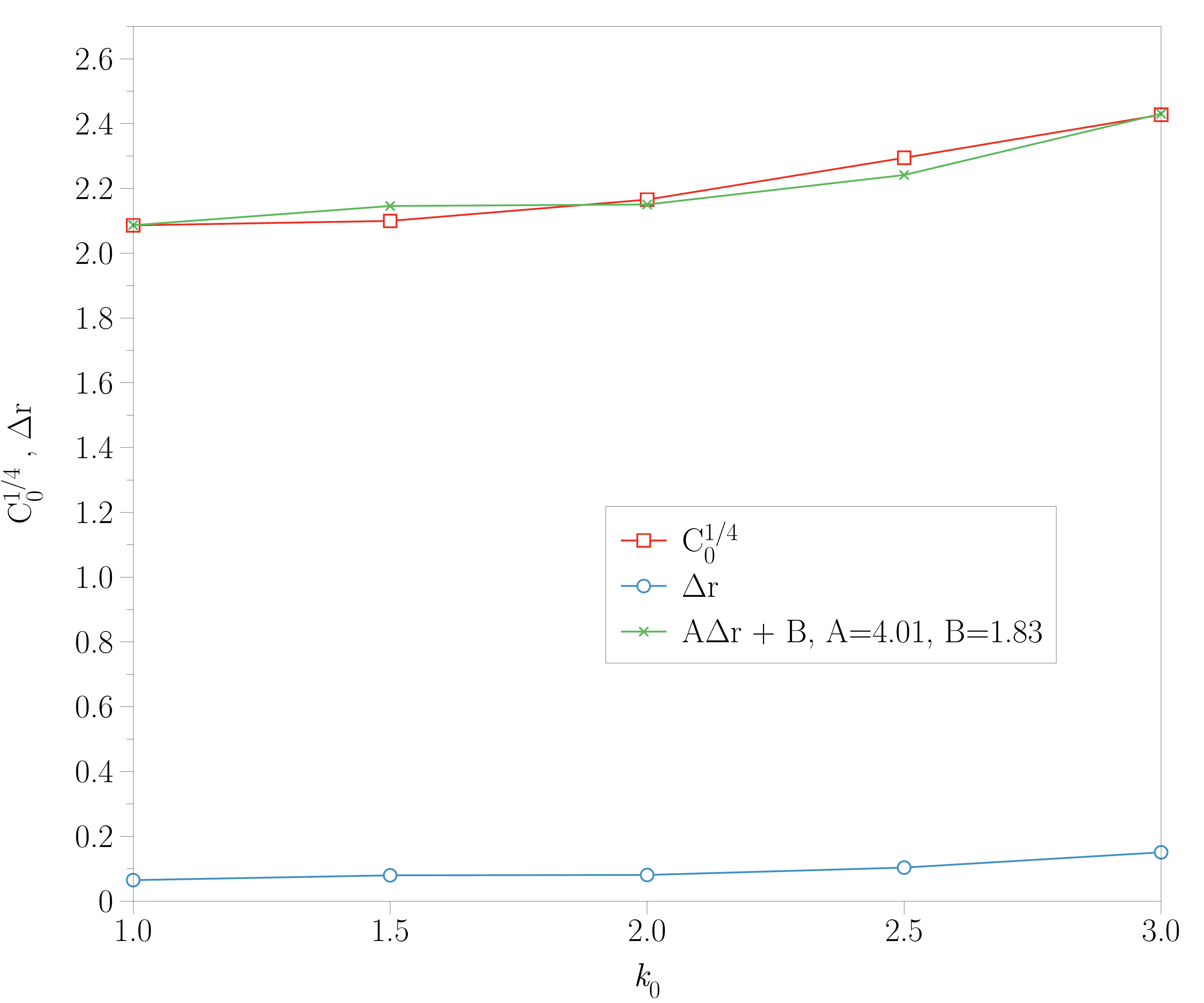}}
\caption{The zero mode fluctuation ($C_0^{1/4}$) compared to the difference in the sampled radius from the actual radius due to lattice effects ($\Delta r$). 
\label{fig:zeromode}}
\end{figure}

Results of this calculation are shown in Figure~\ref{fig:zeromode}. Without any modification, the values for $\Delta r$ that we measured are smaller than the $C_0$s but they show similar behavior. The fit can be made better by multiplying and adding a constant to $\Delta r$. These constants arise from the fact that the assumptions we made in deriving (\ref{eq:c0deltar}) -- constant variance and perfect sphericity -- were not completely correct using our data set (there are small but non zero modes $>$ 0 in our simulations). The results in Figure \ref{fig:zeromode} are an indication that the zero mode fluctuations are indeed induced by the CDT lattice.

\subsection{Small $C_l$ for $l>0$} \label{lgt0}

The power in $l$'s greater than zero is small in all cases, but it is still non-zero. If the spacetimes were truly spherical, with no fluctuations, all of these values would be zero. Is there a small but interesting dynamic in the theory itself that causes these modes to be active? Within the context of this work it is unclear where these modes come from. It is plausible that they arise from lattice effects. If so, then as the lattice spacing decreases, theses modes would also decrease. Decreasing lattice spacing requires an increase in the initialization volume, though, and generating enough sweeps to calculate good statistics is problematic at high volume using available hardware. Studying these small, high $l$ modes is therefore left to future work.

\section {Conclusions}  \label{conclusion}

As Figures \ref{fig:powers} and \ref{fig:fluctpowers} and the analysis in Section \ref{results} suggest, the geometry that the CDT simulations produce are very close to being perfectly spherical with no fluctuations. Because reduced phase space quantization leads to a quantum spacetime with no degrees of freedom, these results imply that, at least in (2+1) dimensions, the path integral quantization implemented by CDT is, at least very nearly, equivalent to the canonical approach. It remains to be seen if this result can be extended to other 2+1 dimensional topologies. In CDT we have, for the first time, a tool to numerically explore these fascinating theories and perhaps cast some light on the (3+1)-dimensional case.

\subsection*{Acknowledgments}

This work was supported in part by the Department of Energy under grant DE-FG02-91ER40674. Thanks to Steve Carlip and Rajesh Kommu for all of their helpful insights.

\bibstyle{unsrt}
\bibliography{Testing_Lattice_Quantum_Gravity_in_2+1_Dimensions-Michael_Sachs}

\end{document}